\documentclass{aa} 

\usepackage{graphicx}
\usepackage{txfonts}
\usepackage{multirow}
\usepackage{pythonhighlight}
\usepackage{longtable}
\usepackage{lscape}
\usepackage{hyperref}
\usepackage{listings}

\definecolor{codegreen}{rgb}{0,0.6,0}
\definecolor{codegray}{rgb}{0.5,0.5,0.5}
\definecolor{codepurple}{rgb}{0.58,0,0.82}
\definecolor{backcolour}{rgb}{0.95,0.95,0.92}

\lstdefinestyle{mystyle}{
  backgroundcolor=\color{backcolour}, commentstyle=\color{codegreen},
  keywordstyle=\color{magenta},
  numberstyle=\tiny\color{codegray},
  stringstyle=\color{codepurple},
  basicstyle=\ttfamily\footnotesize,
  breakatwhitespace=false,         
  breaklines=true,                 
  captionpos=b,                    
  keepspaces=true,                                    
  numbersep=5pt,                  
  showspaces=false,                
  showstringspaces=false,
  showtabs=false,                  
  tabsize=2
}
\begin{document} 

   \title{Dynamics of tidal tails of open clusters: I. effects of bar, spiral arms and giant molecular clouds}

   \subtitle{}

   \author{Janez Kos\inst{1}
          \and
          Jovana Risojevi\'{c}\inst{1}
          \and
          Samo Ilc\inst{1}
    }

   \institute{Faculty of mathematics and physics, University of Ljubljana, Jadranska 19, 1000 Ljubljana, Slovenia\\
              \email{janez.kos@fmf.uni-lj.si}
}

   \date{Received September 15, 1996; accepted March 16, 1997}

 \abstract
   {Open clusters gradually dissolve, and their stars disperse into the Galactic field. Lost stars form tidal tails--elongated streams that trace the cluster's orbit ahead of and behind its core. From the shape and orientation of the tidal tails, it is possible to infer the shape of the gravitational potential governing the cluster's motion. The orbits of open clusters, including those in the Solar neighbourhood, are sensitive to the gravitational potential of the inner Galaxy, which is dominated by the Galactic bar.}
   {Using n-body simulations of synthetic and real open clusters, we investigate how sensitive the shapes and orientations of tidal tails are to variations of the gravitational potential of the Milky Way. We consider the effects of the bar as well as spiral arms, giant molecular clouds (GMCs) and satellite galaxies.} 
   {We analyse the stellar distributions within tidal tails using statistical metrics that quantify the differences between tail morphologies. Such non-parametric approach enables us to efficiently explore tidal tails across a large parameter space of gravitational potential models.}
   {We find that the Galactic bar--particularly its pattern speed--has a strong influence on the orbits of open clusters and the shape of their tidal tails. Spiral arms have a limited effect, and satellite galaxies do not significantly disturb the tidal tails of nearby open clusters. Perturbations by GMCs affect most clusters, with distortions stronger than those by the bar observed in old and in-plane clusters. We identify nearby open clusters that are most sensitive to the pattern speed of the bar. }
   {By observing the tidal tails of a handful of well-selected nearby clusters, we should be able to measure the pattern speed of the bar with a precision in the order of $1\ \mathrm{km\,s^{-1}\,kpc^{-1}}$ independently from length and orientation of the bar. We will present the observability of tidal tails in the second paper.}


   \maketitle
%
\section{Introduction}

Open star clusters survive only briefly as gravitationally bound or spatially coherent structures--on average, for roughly one orbit around the Galaxy \citep{krumholz19}.  They begin losing stars during star formation phase \citep{stoop24, fajrin25}, and gradually dissolve under the influence of internal kinematics and external perturbations. Due to the conservation of angular momentum, stars that have evaporated or been ejected from a cluster occupy orbits that closely follow that of the cluster itself. Stars leading the cluster move slightly inward, while trailing stars move outward relative to the cluster's orbit, forming characteristic S-shaped tidal tails \citep{kupper08, kupper10, kos24}. Tidal tails can extend several kiloparsecs along the cluster orbits, and are as such great tracers of the shape of the orbit. Knowing the shape of the orbit from observations of tidal tails of globular clusters and satellite galaxies (also called stellar streams) has been used extensively to study the gravitational potential of the Milky Way and its dark matter halo \citep{callingham19, posti19, bonaca25}. The same concept applies to open clusters, whose tidal tails could be used to study the gravitational potential of the inner Milky Way.

The gravitational potential of the inner Milky Way is characterised by a bar. An elongated rotating bar produces a highly non-axially symmetric potential that affects the shape of orbits even beyond the Solar galactocentric distance \citep{kim25}. In the literature, the bar pattern speed measurements range from around $\Omega_b=25\ \mathrm{km\,s^{-1}\,kpc^{-1}}$ \citep{horta25} to $\Omega_b=35\ \mathrm{km\,s^{-1}\,kpc^{-1}}$ \citep{chiba21, clarke22, dillamore24} to $\Omega_b=55\ \mathrm{km\,s^{-1}\,kpc^{-1}}$ \citep{monari16b, antoja14, debattista02}, and the bar length estimates vary from $5\ \mathrm{kpc}$ to $3\ \mathrm{kpc}$, respectively. Hence the bar is likely long and slow or fast and short, the former being preferred in the most recent literature. The degeneracy between the length and pattern speed \citep{gerhard11} arises from observational limitations, definition of the bar length \citep{lucey23} and coupling of the bar with the spiral arms pattern and the disk \citep{monari16, michtchenko18}. 

The most precise measurements of the bar pattern speed are obtained by combining stellar kinematics in the bar region \citep{sanders19}, gas dynamics \citep{li22}, and kinematic signatures in the Solar neighbourhood \citep{lucchini24}. The same resonances that affect the kinematics of stars in the Solar neighbourhood also perturb the orbits of clusters and drive the shape of their tidal tails; however, from the tidal tails we can infer the shape of the cluster orbits, and consequently the gravitational potential. Such an additional method of measuring the pattern speed can improve accuracy, and break the degeneracies between the pattern speed and other parameters (position angle, bar length) that appear in most methods. Clusters on eccentric orbits are most sensitive to the potential of the bar close to the pericentre. When a cluster aligns with the major axis of the bar close to the pericentre, this changes the cluster orbit, even if the cluster is not in an orbital resonance with the bar. Because of the precise timing of such an alignment, and the fact that different clusters passed their pericentre at different times, the measurement of the pattern speed using tidal tails has little degeneracy with other measured parameters, such as the orientation of the bar. 

Even clusters that have a pericentre close to the Solar galactocentric radius are sensitive to the corotation resonance (CR) or the outer Lindblad resonance (OLR). Bar's corotation resonance happens within 1 kpc of the Solar galactocentric radius if the bar has a pattern speed between $25\ \mathrm{km\,s^{-1}\,kpc^{-1}}$ and $33\ \mathrm{km\,s^{-1}\,kpc^{-1}}$. These values are very close to the pattern speed of the bar found in recent literature and we expect that some clusters within $1\ \mathrm{kpc}$ from the Sun, such as those studied in this work, are affected by the CR. If the bar pattern speed is larger than $\sim40\ \mathrm{km\,s^{-1}\,kpc^{-1}}$, the outer Lindblad resonance would affect cluster within 1 kpc from the Sun instead of the CR. We note that exact limits also depend on the eccentricity of the cluster orbits and the current position of the cluster on the orbit. It appears that regardless of what is the true pattern speed of the bar and the true gravitational potential of the Milky Way, we should be able to observe many clusters within $\sim1.5\ \mathrm{kpc}$ from the Sun that are close to one of the resonances -- consequently finding clusters with tidal tails that are highly sensitive to the parameters of the bar.  
The main caveat of using the shapes of tidal tails to infer the underlying gravitational potentials is the difficulty of finding stars in the tidal tails of open clusters. We have shown that stars far from the cluster cores can be found if we assume a gravitational potential of the Galaxy and use simulations of cluster dissolution and stellar densities in the Galaxy to constrain the parameter space in which the escaped stars can be found \citep{kos24}. Trying to use such observations to constrain the parameters of the Galactic gravitational potential, such as the mass and pattern speed of the bar, leads to a circular argument, as the potential has already been used to find the stars in the first place. This fallacy is clearly illustrated in \citet{thomas23}. 

Stars in tidal tails close to the cluster core have also been found in many recent studies without using the priors from the simulations \citep{meingast21, tarricq22, bhattacharya22, hunt23, risbud25}. We expect that finding stars far in the tidal tails with non-informative priors will become possible in the near future, which will present the opportunity to actually measure the gravitational potential of the Milky Way from the shape of the tidal tails of open clusters. We emphasise here that all current studies that connect the Milky Way gravitational potential with the tidal tails have to do the opposite: assume a gravitational potential to find the shape of the tidal tails. With the arrival of the fourth data release by Gaia and first data releases by major new spectroscopic surveys by the end of 2026 or in 2027, we soon expect an increase in precision of distances and proper motion measurements, and an immense increase in precise radial velocity measurements below Gaia's magnitude limit \citep{dejong19, jin24, kollmeier26}. We also expect advancements in models of the Milky Way's stellar populations \citep{robin22}, extinction maps \citep{lallement22}, and selection functions \citep{cantat23}. This will allow us to find the stars in tidal tails further from the cluster cores than currently possible. In this work, we show that finding reliable cluster members just a couple hundred parsecs from the core would be enough to infer the pattern speeds of the bar with a precision close to $2\ \mathrm{km\,s^{-1}\,kpc^{-1}}$. 

The shape of the tidal tails is further influenced by interactions with spiral arms \citep{gieles07} and with GMCs \citep{gieles06}. We attributed the observed asymmetries of tidal tails in \citet{kos24} to interactions with GMCs. GMCs were also discussed as the most probable reason for well studied highly asymmetric tidal tails of the Hyades \citep{jerabkova21} and for the existence of young coreless streams in the Milky Way \citep{miller25}. In this work we simulated the development of tidal tails in a model of the Milky Way gravitational potential that includes thousands of GMCs with a total mass of $10^9\ M_\odot$. We found that the disruption of tidal tails by GMCs is frequent, and could be confused with the tidal effects of the bar in some cases. 

In this work, we explore in-depth the sensitivity of tidal tails of open clusters to the parameters of the bar, particularly the pattern speed. The work presented in this paper is based on n-body simulations of synthetic and real open clusters. In a following paper, we will study the observability of tidal tails in the data of Gaia DR3 and DR4, with the inclusion of radial velocities from various spectroscopic surveys. 

The paper is structured as follows. In Section \ref{sec:potential} we present the gravitational potential of the Milky Way that we use throughout this work, together with the potentials of spiral arms, satellite galaxies, galactic bar, and GMCs. In Section \ref{sec:simulation} we describe how the n-body simulations were performed in this work. In Section \ref{sec:sensitivity} we present the analysis of the simulations of synthetic clusters that most systematically demonstrate the sensitivity of the shape and position of tidal tails on perturbations induced by the perturbations in the gravitational potential of the Milky Way. Section \ref{sec:real_clusters} repeats some of the analysis from the previous section, but using simulations of real open clusters currently found within 1 kpc from the Sun. We conclude with Section \ref{sec:conclusions}. The paper includes an extensive appendix, where code and formulae are shown that can be used to reproduce our gravitational potentials. In the appendix, we also discuss statistical metrics that we used to evaluate and compare the variations of shapes of tidal tails and structures within them. 

\section{Gravitational potential of the Milky Way}
\label{sec:potential}

Our n-body code can compute the motion of stars in some external potential. Decoupling the forces between stars in a cluster from external forces improves the computational time, because the external potential is smooth and the external forces on each individual star can be computed much less frequently than the forces inside the cluster. The performance of computations can be further increased if the external forces can be computed analytically. Hence, we represent the external forces (i.e., the gravitational potential of the Milky Way) with a sum of simple analytical potentials readily available in \textsc{Galpy} \citep{bovy15} \footnote{\url{http://github.com/jobovy/galpy}}. We use (almost) the same gravitational potential of the Milky Way throughout this work and refer to it as the common gravitational potential. 

\begin{figure}[!ht]
    \includegraphics[width=\columnwidth]{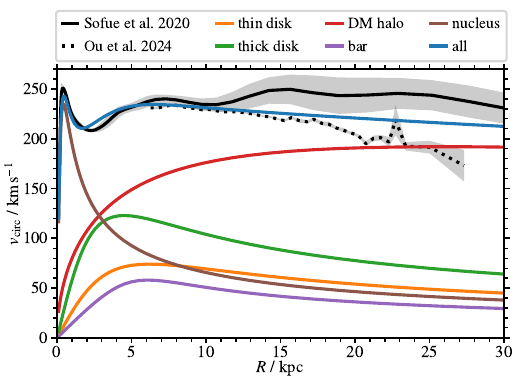}
    \caption{Rotational curve of the common gravitational potential (blue line). Rotational curves of individual components that constitute the common potential are also shown. Rotational curves from \cite{sofue20} and \cite{ou24} are shown in black.}
    \label{fig:rotational_curve}
\end{figure}

To construct the common gravitational potential, we use literature values for the masses and sizes of the nucleus, bulge, bar, thin and thick disks, and the halo. Because these values can vary significantly in the literature, we aimed to produce a common gravitational potential with a rotational curve that matches the observations in \citet{sofue20} and a rotational velocity at Solar galactocentric radius of $v_{\mathrm{circ}, R_\odot}=233.4\ \mathrm{km\,s^{-1}}$ \citep{drimmel18}. The rotational velocity of our model and its components is shown in Figure \ref{fig:rotational_curve}. We were unable to reproduce the bump seen in the observed rotational curve at $R_\mathrm{GC}>12\ \mathrm{kpc}$; however, many authors observe a lower rotational curve in this region \citep[e.g.][also shown in Figure \ref{fig:rotational_curve}]{ou24}.

The components from which we built the potential are described in more detail in Appendix \ref{sec:potential_appendix}. Notably, our potential includes a thick disk with twice the mass of the thin disk. This is in accordance with several modern studies \citep[e.g.][]{khanna25, xiang25}, which, in contrast to older literature, find that the thick disk dominates the surface density at Solar $R_\mathrm{GC}$. See also a review by \citet{hunt25}. 

\subsection{Variations and perturbations}

\begin{figure*}[!ht]
    \centering
    \includegraphics[width=\textwidth]{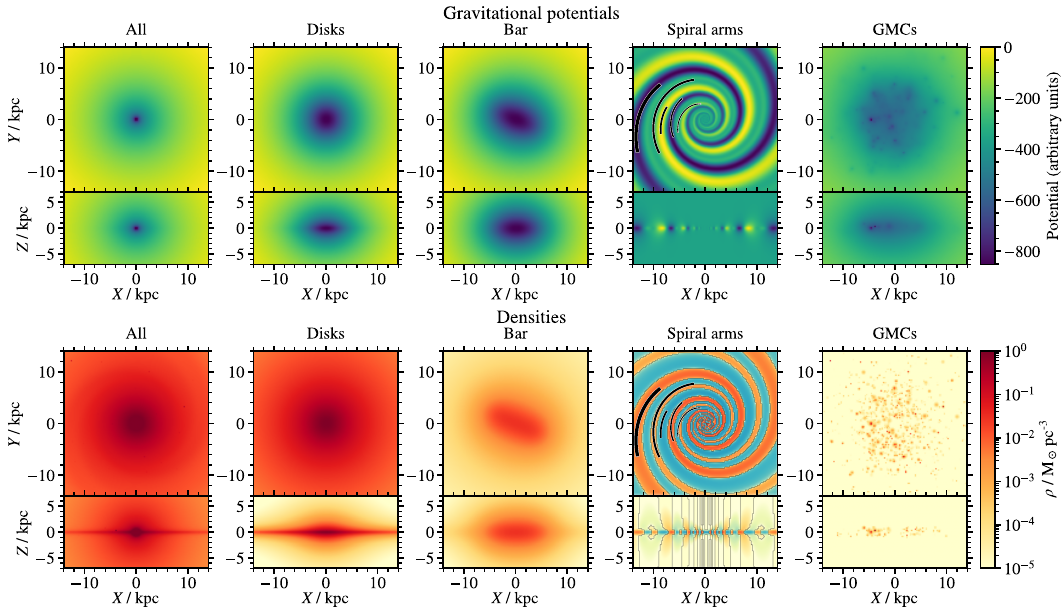}
    \caption{Potentials used in this work. Left: the complete potential including spiral arms and GMCs. Top panels show the potential in the $Z=0$ plane, and in the $Y=0$ plane. Below, we plot the density corresponding to the above potential. Following from left to right: potential and density of disks (thin and thick), the bar, the spiral arms, and the GMCs. Spiral arms are implemented as a perturbation of the common potential by adding density in the arms and reducing the density in between the arms. Negative density is shown with blue tones. Positions of GMCs are random and would be different in each simulation. Potentials shown in each of the top five panels are plotted with different scales. Spiral arms from \citet{reid14} are marked with black lines.}
    \label{fig:potentials}
\end{figure*}

We studied the development of tidal tails in response to variations of the external gravitational potential by using the common gravitational potential, varying only one or two parameters at a time. The possible variations, as well as the additional components we added to the common potential, are described below. The common potential, disks, bar, spiral arms and GMCs are illustrated in Figure \ref{fig:potentials}.

\subsubsection{Bar}

We included a simple bar into our common gravitational potential, whose parameters we can easily vary in the tests and analyses of clusters presented in this work. We realise that the gravitational potential of our bar is simplified and that more realistic models exist \citep[e.g.][]{khalil25}. A simple model such as ours enabled us to compute external forces in n-body simulations analytically \citep{long92}. Even a simple potential has enough free parameters (mass, length, pattern speed, orientation) to allow for a comprehensive analysis of the effect of the bar on the tidal tails of open clusters. We included a bar in simulations of all other perturbations (GMCs, spiral arms, satellite galaxies), because we expected the bar to have a dominant effect and because the secondary perturbations should also be studied in a realistic, non-axisymmetric potential.

\subsubsection{Spiral arms}
\label{sec:spirals_pot}

The shape, prominence, and dynamics of the Milky Way's spiral arms are highly debated topics in modern literature and have been for decades \citep{oort58}. Observations of masers have revealed a four-armed spiral structure \citep{bobylev14, hou14}, while older observations of stellar densities and dust structures favoured a two-armed spiral \citep{drimmel00, churchwell09}. The four-armed spiral has been corroborated by the observations of Cepheides by the Gaia mission \citep{Lemasle22, drimmel25}, young stars \citep{poggio21}, open clusters \citep{castroginard21}, and H\textsc{ii} regions \citep{reid19, shen25}. A four-arm structure is also favoured by observations of molecular gas \citep{englmaier11}. However, an agreement on the shape of the spiral pattern has not been reached, not even in the Solar neighbourhood \citep[see comparison plots in ][]{drimmel25, shen25,poggio21}.

The surface density amplitude of spiral arms above the mean background density varies in the literature from around 10\% \citep{eilers20} to 20\% \citep{widmark24} at the Solar $R_\mathrm{GC}$. It is also not clear which arms should have a higher density amplitude and whether their prominence is consistent across the whole spiral pattern. Discussions on these issues can be found in the papers cited above. 

Particularly, observations of open clusters have recently shown that the pattern speed of the spiral structure is not constant (as it would be for a density wave), but is different for each, or at least most arms \citep{castroginard21, liu25}. It varies between 12 and 50 $\mathrm{km\,s^{-1}\,kpc^{-1}}$.

Without clear conclusions on a global shape of the spiral arms, we used a highly simplified potential to describe the perturbation by spiral arms in this work. We acknowledge that better models of the potential of the disk with spiral arms exist \citep{khalil25}; however, our aim is to test whether spiral arms affect the shape of open clusters, for which a simple model is sufficient. A simple model reproducible in \textsc{Galpy} also allows for faster integration of orbits. 

We constructed a gravitational potential using \textsc{Galpy}'s \texttt{potential.SpiralArmsPotential()} function with parameters for four spiral arms made of two pairs. Each pair of arms is described with a concentrated arms model \citep{cox02}. The first pair describes the dominating arms, and the second pair describes weaker arms with a density amplitude $1/2$ of that of the dominating pair at Solar $R_\mathrm{GC}$. All four arms have a pitch angle of $12.6^\circ$. The reference angle of the dominating pair at Solar $R_\mathrm{GC}$ is $127.8^\circ$ and for the weaker pair $52.2^\circ$. The scale height of the spiral arm perturbation is the same as the scale height of the thin disk, and the radial scale length is twice the scale length of the thin disk, so the perturbation can persist throughout the outer disk. The density profile of our spiral arms' potential aligns with the four major spirals in \citet{reid14}, as illustrated in Figure \ref{fig:potentials}. The above definitions hold at the present time, and we can assign the spiral arms some pattern speed. In our analysis, the pattern speed was constant, and there was no differential rotation of individual arms. 

\subsubsection{Giant molecular clouds}

We modelled the GMCs with Plummer potentials that describe a smooth distribution of mass with a characteristic radius:
\begin{equation}
    \Phi_\mathrm{GMC}(r;M_\mathrm{GMC},a)=-\frac{GM_\mathrm{GMC}}{\sqrt{r^2+a^2}},
\end{equation}
where $a=25\ \mathrm{pc}$ is a fixed Plummer radius, $M_\mathrm{GMC}$ is the mass of a GMC, and $r$ in the above formula is the distance from the centre of each GMC. Masses of GMCs  have a truncated power-law distribution:
\begin{equation}
    \frac{dN}{dM}\propto M^{-\alpha};\quad M_\mathrm{min}<M<M_\mathrm{max},
\end{equation}
with a cumulative distribution function (CDF) \citep{colombo14}:
\begin{equation}
    N(M'>M)=N_0\left[ \left(\frac{M}{M_\mathrm{max}}\right)^{1-\alpha} -1 \right].
    \label{eq:cdf}
\end{equation}
Note that this CDF counts the number of GMCs with masses greater than a given value. From Equation \ref{eq:cdf}, we can derive the total mass of the GMCs or the number of GMCs needed to reach some total mass. For the power-law index, we use $\alpha=1.72$, which is the mean index for the distribution of GMC masses in the Milky Way reported by \citet{rice16}. We truncated the power-law to masses between $10^4\ M_\odot<M_\mathrm{GMC}<10^7\ M_\odot$. In our simulations, we only varied the total mass of the GMCs.

We positioned the GMCs on orbits with initial velocity $(U, V, W)=(0, v_\mathrm{circ},0)$; this is, on orbits as close to circular as possible, with $v_\mathrm{circ}^2=R(\partial\Phi/\partial R)$. To model the spatial distribution of GMCs' initial positions, we constructed a `gas disk' with a Miyamoto Nagai density profile with scale length $R_\mathrm{GMC}=2.0\ \mathrm{kpc}$ and scale height $h_\mathrm{GMC}=65\ \mathrm{pc}$ \citep{miville17}. The spatial distribution of GMCs was then constructed by randomly sampling this density profile, except for the central region with $R_\mathrm{GC}<5.5\ \mathrm{kpc}$, where literature suggests a more complicated and generally flatter distribution \citep{miville17}. Because the orbits of clusters used in this work never reach such a small $R_\mathrm{GC}$, we modelled the central region with a uniform surface density. Such sampling produced a radial distribution of GMCs very similar to that in \citet{miville17}. Potentials of the GMCs were then added to the common gravitational potential (the `gas disk' was discarded at this point, because it is represented by the GMCs). We reduced the mass of the thin disk so the rotational velocity at $R_\odot$ remained the same as before we added the GMCs.

Including GMCs into the external potential implemented in \textsc{Galpy} is computationally ineffective. Each additional component that constitutes an external potential requires an additional term to find the force on each particle in an n-body simulation. To populate the Galaxy with a sufficient mass of GMCs, thousands of GMCs would have to be included in the common potential (4349 GMCs are shown in Figure \ref{fig:potentials}). This can be mitigated by only adding GMCs with orbits between the apocenter and pericenter of the cluster's orbit. This reduces the number of needed GMCs to around 300 for highly eccentric orbits ($e=0.2$) and less than 100 for orbits similar to Solar.

\subsubsection{Satellite galaxies}

Large Magellanic Cloud (LMC), Small Magellanic Cloud (SMC) and the Sagittarius dwarf galaxy (SDG) are the three most massive satellite galaxies of the Milky Way \citep{mcconnachie12}. We assumed that they have a detectable effect on the orbits of open clusters. We tested this by adding their potentials to the common gravitational potential and compared the orbits and tidal tails with or without the additional galaxies in the external potential. After verifying that LMC, SMC, and SDG have a minor effect on the orbits, we decided not to include any other satellites in this study. They are less massive and further from the inner Galaxy than the SDG (with a pericenter of 15 kpc in the potential of the Milky Way used in this work\footnote{Our pericentre matches with values in other simulations \citep[e.g.][]{law10} and with the pericentre of the SDG tidal stream \citep{ramos20}.}) and should have negligible effects. 

Table \ref{tab:satellites} lists the present parameters of all three galaxies. We obtained them by \textsc{Galpy} function \texttt{Orbit.from\_name()}, which queries the \textsc{Simbad} database. The orbits were initialised from these parameters, integrated back in time for the age of each cluster, and added into the common potential before the n-body simulation was started. The potential of each satellite galaxy is represented by a Plummer potential with a characteristic radius of $3\ \mathrm{kpc}$ and masses $M_\mathrm{LMC}=5.0\, 10^{10}\ M_\odot$ \citep{warren25}, $M_\mathrm{SMC}=7.0\, 10^{9}\ M_\odot$ \citep[this is $\sim1/7\, M_\mathrm{LMC}$][]{bekki09, pardy18}, $M_\mathrm{SDG}=4.0\, 10^{8}\ M_\odot$ \citep{vasiliev20}. The shape and radius of the potential are mostly irrelevant, because no open cluster approaches any satellite galaxy to within several kpc. 

\begingroup
\begin{table*}[!ht]
    \caption{Parameters that we used to initialise the orbits of satellite galaxies. Cartesian positions and velocities were computed from celestial coordinates and velocities.}
    \setlength{\tabcolsep}{3.13pt}
    \renewcommand{\arraystretch}{1.3}
    \begin{tabular}{l c c c c c c c c c c c c}
         Name & $\alpha$ (ICRS) & $\delta$ (ICRS) & $d$ & $\mu_\alpha \cos \delta$ & $\mu_\delta$ & $v_\mathrm{r}$ & $X$ & $Y$ & $Z$ & $U$ & $V$ & $W$\\\hline
          & deg. & deg. & kpc & $\mathrm{mas\,yr^{-1}}$ & $\mathrm{mas\,yr^{-1}}$ & $\mathrm{km\,s^{-1}}$ & kpc & kpc & kpc & $\mathrm{km\,s^{-1}}$ & $\mathrm{km\,s^{-1}}$ & $\mathrm{km\,s^{-1}}$\\ \hline\hline
         LMC & 78.77$^{1}$ & -69.01$^{1}$ & 50.1$^{2}$ & 1.85$^{3}$ & 0.234$^{3}$ & 262.2$^{4}$ & -1.1281 & -41.0436 & -27.8452 & -55.3347 & -222.9824 & 209.4193\\
         SMC & 16.26$^{1}$ & -72.42$^{1}$ & 62.8$^{5}$ & 0.797$^{3}$ & -1.22$^{3}$ & 145.6$^{4}$ & 15.0926 & -38.0847 & -44.1887 & 28.1002 & -179.9001 & 175.4862\\
         SDG & 283.8292$^{6}$ & -30.5453$^{6}$ & 26$^{5}$ & -2.65$^{7}$ & -0.88$^{7}$ & 140.0$^{4}$ & 16.9522 & 2.4461 & -6.4059 & 233.3132 & 36.6963 & 221.8381\\\hline
    \end{tabular}
    \setlength{\tabcolsep}{4pt}
    \begin{tabular}{l l l l}
         $^{1}$ \citet{paturel03} & $^{3}$ \citet{kallivayalil13} & $^{5}$ \citet{drlica20} & $^{7}$ \citet{ibata97}\\
        $^{2}$ \citet{bono10} & $^{4}$ \citet{mcconnachie12} & $^{6}$ \citet{carlsten21} & \\
    \end{tabular}

    \label{tab:satellites}
\end{table*}
\endgroup

\section{Simulating formation of tidal tails}
\label{sec:simulation}

\subsection{The coordinate system}

\begin{figure}[!ht]
    \centering
    \includegraphics[width=0.92\columnwidth]{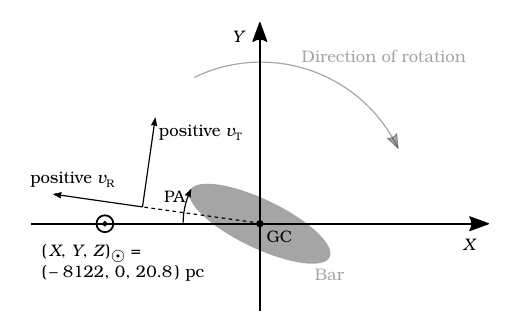}
    \caption{Coordinate system used in this work. $Z$ axis (not shown) points out of the drawing. $PA$ marks the position angle of the bar.}
    \label{fig:coordinate_system}
\end{figure}

In this work, we use a Cartesian galactocentric coordinate system in which the Sun is at a position $(X, Y, Z)=(-8122.0, 0.0, 20.8)\ \mathrm{pc}$ \citep{gravity18, bennett19}. The local standard of rest (LSR) at $X=-8122.0\ \mathrm{pc}$ has a velocity $(U, V, W)=(0.0, 233.4, 0.0)\ \mathrm{km\,s^{-1}}$ \citep{drimmel18}, making the Galaxy rotate in the positive $Y$ direction. In this frame, the Sun has a velocity $(U, V, W)_\mathrm{LSR}=(12.9, 12.2, 7.78)\ \mathrm{km\,s^{-1}}$ \citep{drimmel18}. We also use a cylindrical coordinate system in which positive velocities $v_\mathrm{R}$ and $v_\mathrm{T}$ point in the direction out of the Galactic centre and in the direction of rotation, respectively. The coordinate system is illustrated in Figure \ref{fig:coordinate_system}.

\subsection{Initial conditions}

The evolution of all our gravitational potentials can be reversed in time. This means that the orbit of the cluster centre can be integrated back in time to the cluster's age, and then evolved forward in time, and the cluster centre must end up at the same current position. We can use the current position and velocity of the cluster centre as boundary conditions from which we can compute the initial condition for cluster centre at the cluster's age. Due to the limited precision of the n-body simulations, the dynamical evolution of the stars in the cluster is in practice not reversible in time. Thus, the n-body simulation can only be initiated at the age of the cluster at the position where the cluster centre was at the time. All our initial conditions were constructed in such a way. 

We considered the initial position of the cluster centre as its centre-of-mass and populated the cluster with stars. Instead of the cluster centre, in the forward n-body simulation, we only evolved the positions and velocities of individual stars until the current time. We initialised all our simulated clusters with a King 3D density profile with a dimensionless central potential $W_0=7.0$ \citep{king62} and a fixed number of stars, whose masses sampled a Kroupa mass function \citep{kroupa01}. In all simulations, we populated the cluster with enough stars that at least half remained in the cluster core at the end of the simulation. Initial stellar masses were limited between $0.1\ M_\odot<M<125\ M_\odot$. We fixed the virial radius to $R_\mathrm{vir}=2.2\ \mathrm{pc}$ and set the velocities of stars such that clusters were virialised. Metallicity \citep[$Z=0.02$,][]{pols98}, stellar radii, luminosities, and other stellar parameters irrelevant to our study were set by the stellar evolution model used in \textsc{SeBa} \citep{portegies96, zwart12}. For simulations used in this work, we produced initial conditions with no binaries; however, we experimented with simulations that include binaries \citep{mcmillan12, wang20}. Because in this work we aimed to simulate clusters across a large grid of different potentials,  computational time was a major concern. In the next paper, we aim to explore the observability of tidal tails in real data; therefore, simulations of realistic binaries will be needed.

To define the centre of the cluster during the forward simulation or in the final snapshot, we can use either the centre-of-mass or the maximum star density. The centre-of-mass should follow the orbit we computed by integrating backwards in time, and the centre defined by a maximum density matches the definition of the cluster centre used in observational papers \citep[e.g.][]{hunt23}. In practice, integrating orbits back and then forward again does not necessarily return the cluster's centre to the same position due to numerical precision and stellar evolution, which change the masses of stars. In our tests, the largest error was accumulated when we added GMCs to the gravitational potential. The clusters typically finished the simulations up to $20\ \mathrm{pc}$ from their current positions when we use the maximal density to define the centre of the cluster. The error for the centre-of-mass is around half that. The error in simpler potentials (for example just in the common potential) is around $5\ \mathrm{pc}$. To mitigate the problem of clusters not ending the simulation at the correct positions, we artificially shifted the clusters, along with all their stars, to the correct positions (positions and velocities we called the boundary condition) at the end of the simulation. 

\subsection{N-body simulations}

From the initial conditions, we computed the cluster's evolution to the current time using an n-body simulation, accounting for the evolution of the external gravitational potential and stellar evolution. The computation was performed within the AMUSE framework\footnote{\url{https://github.com/amusecode/amuse}} \citep{webb20, portegies09, portegies13, pelupessy13, portegies18}, which traced the position of the stars, evolution of the external gravitational potential, and communicated with a stellar evolution code. The gravitational potential of the Galaxy and its components was built with the \textsc{Galpy} package, which can communicate with AMUSE \citep{webb20} and report the forces from the external gravitational potential acting on each star at any time in the simulation. Forces between stars were computed by a fourth-order Hermite direct n-body integrator ph4 \citep{mcmillan12}, which is part of AMUSE. To speed up the computations, we introduced softening of the star particles of $\epsilon=0.03\ \mathrm{pc}$, which is more than five times smaller than a typical nearest-neighbour distance in the core of a cluster initialised with 2000 stars. This suppresses the creation of close binary systems which would slow down the simulations. To control the precision, we set the \texttt{timestep\_paramater} in the ph4 code to 0.03. When testing, we found that the simulation over $500\ \mathrm{Myr}$ produces results that are statistically indistinguishable for timestep values below 0.06. 

We accounted for stellar evolution, which in our simulation causes mass loss via supernova explosions and stellar winds and accelerates the dissolution of clusters by reducing their total mass. It also causes a notable difference between the cluster's centre-of-mass and peak density at the end of the simulation. We implemented stellar evolution using a \textsc{SeBa} code \citep{portegies96, zwart12}. It models stellar evolution using precomputed prescriptions. It requires an initial mass (computed when we set initial conditions) and an initial metallicity (fixed at Solar metallicity), and can report the mass of each star at each time step of the n-body integrator. It also traces other stellar parameters, such as radius and luminosity, which we did not use in this work. 


\subsection{Distance metrics}

We expect the tidal tails that evolve in different gravitational potentials to be distinguishable at the end of the simulation. The stars in the cluster and in the tidal tails can be seen as points that sample a probability distribution: a likelihood that stars in a particular simulation end up at some position in the Milky Way. This definition can easily be expanded into a likelihood that the stars end up at some position in an arbitrary parameter space; however, in this work, we will only analyse the position of tidal tails in the $XY$ plane of the Milky Way. If two simulations produce tidal tails with different shapes, we should be able to quantify the difference as the distance between two probability distributions. Because we only sample the probability distributions with a small number of stars that end up in the tidal tails, the distance between distributions is not trivial to compute.

We tried several methods for measuring distances between probability distributions to identify the one that best works with the shapes of tidal tails. We were not able to find a single good metric, so we ended up using two metrics--the Kullback-Leibler divergence (KLD) and the Maximum mean discrepancy (MMD).  In Appendix \ref{sec:metrics}, we tested different metrics on a toy model: a simulation of the tidal tails of Hyades that we artificially modified and distorted. We explored which metrics are sensitive to specific features: tail length, tail orientation, core concentration, tail structures, and possible undersampling issues. KLD proved to be extremely sensitive to the orientation of the tails, and MMD to the structures in the tails, such as positions of overdensities. The metrics, and the results of the tests are presented in Appendix \ref{sec:metrics}.

We have decided to use statistical metrics as opposed to describe the tidal tails with morphological parameters (e.g. orientation of the tail, length, position of overdensities). Statistical metrics are more objective and direct measures of differences than morphological parameters, especially in simulations where we can ``observe'' the whole extend of the tails. Statistical metrics compute meaningful distances that can be interpreted in terms of noise and limited number of samples, as opposed to morphological parameters that require arbitrary definitions and thresholds. 

\section{Sensitivity of tidal tails to variations and perturbations}
\label{sec:sensitivity}

\subsection{Spiral arms}

We performed simulations of tidal tail formation using the common Milky Way potential with added spiral arms as described in Section \ref{sec:spirals_pot}. We set the amplitude of spiral arms to either $10\%$ or $20\%$ and the simulation run-times (age of the simulated clusters) to $500\ \mathrm{Myr}$ and $1500\ \mathrm{Myr}$. We varied the pattern speed of the spiral arms between $20\ \mathrm{km\,s^{-1}\,kpc^{-1}}\leq\Omega_\mathrm{s}\leq 36\ \mathrm{km\,s^{-1}\,kpc^{-1}}$. Clusters were always initialised with 2000 stars. All clusters were put on orbits in the plane of the Milky Way, and had the present-day velocities $v_\mathrm{R}$ and $v_\mathrm{T}$ set to one of $-30$, $-15$, $0$, $15$, or $30\ \mathrm{km\,s^{-1}}$, i.e. we simulated the clusters on a grid of 25 different present-day velocities.

\begin{figure*}[!ht]
        \includegraphics[width=\columnwidth]{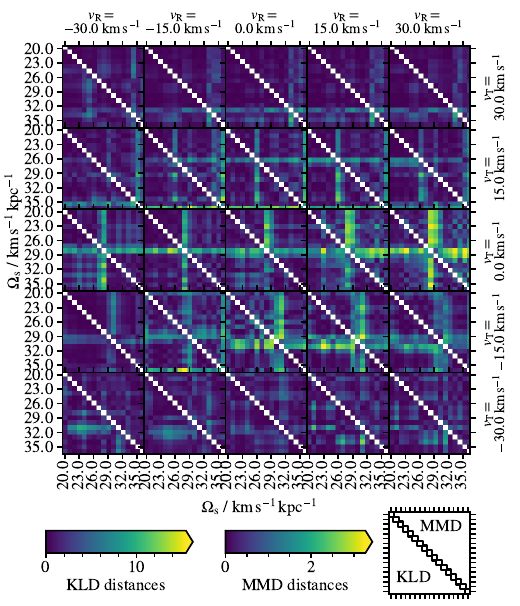} \hfill \includegraphics[width=0.95\columnwidth]{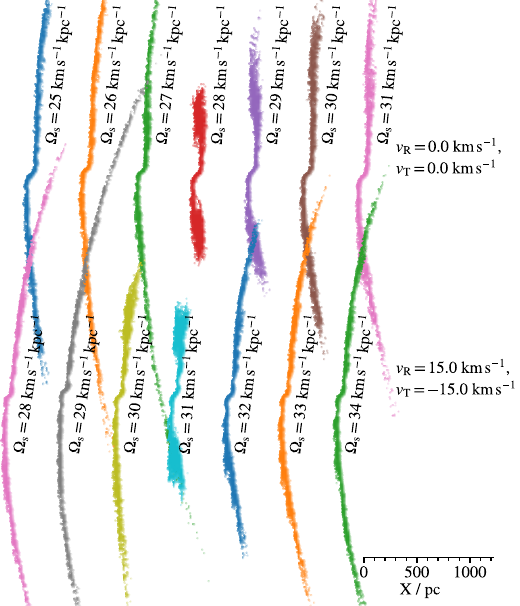}
        \caption{Left: Similarities between all pairs of simulations 1.5 Gyr long with different pattern speed of spiral arms ($\Omega_\mathrm{s}$) and cluster's current $v_\mathrm{R}$ (panels left to right) and $v_\mathrm{T}$ (panels top to bottom). In each panel, we plot the KLD and MMD distances between pairs of simulations with different $\Omega_\mathrm{s}$. Above the diagonals, we plot the MMD distance; below, the KLD distance. We note that the numerical values for MMD and KLD distances are not on the same scale; hence, two colour bars. Right: Shapes of tidal tails close to the CR between clusters and spiral arms for two orbits marked on the right. All clusters are plotted on the same scale, marked at the bottom. }
    \label{fig:spirals_sim}
\end{figure*}

In the $500\ \mathrm{Myr}$ simulation, we detected small effects of the spiral arms on the shape of the clusters close to the CR (between the mean angular velocity of the cluster and the spiral pattern speed). The effects of the spiral arms were negligible or undetected in simulations with a $10\%$ amplitude and minor when we set the amplitude to $20\%$. In the $1500\ \mathrm{Myr}$ simulation with a $20\%$ amplitude, the CR had a strong effect on the shapes of the tidal tails. We evaluated the effects by plotting the distance between the distributions of stars in two simulations with different $\Omega_\mathrm{s}$. A grid of distances for all possible combinations of $\Omega_\mathrm{s}$ then illustrates the similarities between all possible tidal tails in a simulation. Similarities for tidal tails in a simulation for $1500\ \mathrm{Myr}$ and spirals with an amplitude of 20\% are illustrated in Figure \ref{fig:spirals_sim}. Where KLD and MMD distances in Figure \ref{fig:spirals_sim} are large close to the diagonal, the tails are most sensitive to small variations of the pattern speed of the spirals.

When the pattern speed of the spiral arms matches the mean angular velocity of a cluster to within $\sim 1\ \mathrm{km\,s^{-1}\,kpc^{-1}}$, we observe that the most distant parts of the tidal tails are folded toward the cluster centre (Figure \ref{fig:spirals_sim} right). The folded part can overlap with the undisturbed part of the tidal tails or lie very close, almost parallel to it. For the pattern speeds and angular velocities that do not match as well, but do match to within $\sim 5\ \mathrm{km\,s^{-1}\,kpc^{-1}}$, we can only detect a change in the lengths of tidal tails with no clear fold. For reference, the mean angular velocity and eccentricity for clusters in each of the 25 panels can be found in Figure \ref{fig:typical_orbits}.

The CR is strongest for clusters on close to circular orbits, as is evident from Figure \ref{fig:spirals_sim} (left): KLD or MMD distances are largest for clusters on orbits with the smallest eccentricity. We note that the angular velocity can vary significantly along the orbit, yet the CR remains present. For examples illustrated in Figure \ref{fig:spirals_sim} (right), the angular velocities are $\Omega=28.5 \pm 0.8\ \mathrm{km\,s^{-1}\,kpc^{-1}}$, and $\Omega=29.9 \pm 3.5\ \mathrm{km\,s^{-1}\,kpc^{-1}}$ for the top and bottom examples, where the values after the $\pm$ sign are one standard deviations around the mean angular velocity along the orbit of the cluster centre. The conclusion is that the mean angular velocity of the cluster and the pattern speed of the spirals must match extremely well before the CR disturbs the tidal tails. The strongest resonance observed in our simulations is for orbits with $v_\mathrm{R}=30\ \mathrm{km\, s^{-1}}$ and $v_\mathrm{T}=0\ \mathrm{km\, s^{-1}}$. This is most likely the case because this is the only orbit where the mean angular velocity $\Omega=28.0\ \mathrm{km\,s^{-1}\,kpc^{-1}}$ matches exactly the pattern speed, which were computed on a grid with a step of $1\ \mathrm{km\,s^{-1}\,kpc^{-1}}$. 

In our simulations, the CR with spiral arms should cause only minor disturbances to the tidal tails of open clusters. Disturbances are only pronounced when angular velocities of the clusters match the spiral pattern speed extremely well, which is not the case for most clusters. Our simulations show that non-transient spiral arms have to be strong (density amplitude larger than $10\%$) and the perturbations prolonged over $1\ \mathrm{Gyr}$ before the impact of the spirals on tidal tails is the same as the effect of the Galactic bar. We did not simulate more realistic transient spiral arms with non-constant pattern speeds, which likely affect the tidal tails more than spiral arms in our simulation. Such spiral arms can not be studied as systematically as the model used in this work, and we leave this exercise for future studies. 

\subsection{Bar}

To simulate the effects of the bar on the tidal tails, we used the common gravitational potential (with no spirals, GMCs or satellite galaxies) and varied the parameters of the bar. In general, we found that clusters on orbits with the smallest pericentres at around $R_\mathrm{GC}\simeq6\ \mathrm{kpc}$ are most affected by the bar. The effect is substantial for clusters on orbits similar to Solar, and negligible for the clusters with pericentres with $R_\mathrm{GC}>10\ \mathrm{kpc}$. Therefore, in the following analysis, we only discuss the effects of the bar for cluster on orbits similar to Solar or tighter.

\begin{figure*}[!ht]
        \includegraphics[width=0.5\linewidth]{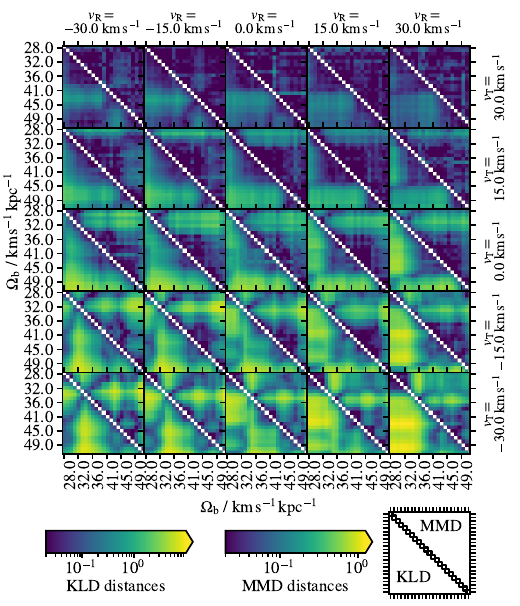} \includegraphics[width=0.5\linewidth]{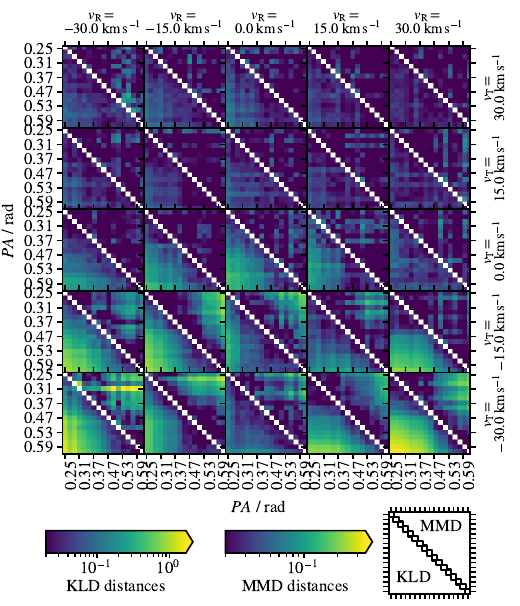}
        \caption{Left: similarities between all pairs of simulations with a varied bar pattern speed $\Omega_\mathrm{b}$ and cluster's current $v_\mathrm{R}$ (panels left to right) and $v_\mathrm{T}$ (panels top to bottom). In each panel, we plot the KLD and MMD distances between pairs of simulations with different $\Omega_\mathrm{b}$. Above the diagonals, we plot the MMD distance; below, the KLD distance. We note that the numerical values for MMD and KLD distances are not on the same scale, hence two colour bars. Right: similarities between all pairs of simulations with a varied position angle of the bar $PA$. 25 panels show the same arrangement of $v_\mathrm{R}$ and $v_\mathrm{T}$ as on the left. }
    \label{fig:bar_sim}
\end{figure*}

In the first simulation, we varied only the bar pattern speed between $28\ \mathrm{km\, s^{-1}\, kpc^{-1}}\leq \Omega_\mathrm{b}\leq 51\ \mathrm{km\, s^{-1}\, kpc^{-1}}$ in steps of $1\ \mathrm{km\, s^{-1}\, kpc^{-1}}$. The run-time was $500\ \mathrm{Myr}$. For each instance of the potential, we computed the initial conditions with present-day velocities on the same grid of 25 different $v_\mathrm{R}$ and $v_\mathrm{T}$ as for the spiral arms simulation, and populated the cluster with 2000 stars. Figure \ref{fig:bar_sim} (left) shows the similarities between all pairs of clusters with varying $\Omega_\mathrm{b}$. The differences between tidal tails get larger with decreasing pericentre (bottom of the plot). For clusters close to the CR and the OLR, both KLD and MMD metrics can distinguish between the shapes of the tails in simulations where $\Omega_\mathrm{b}$ differs by only $2\ \mathrm{km\, s^{-1}\,kpc^{-1}}$. 

Unlike in the case of spirals, the clusters do not have to be as close to the resonances for the tidal tails to be sensitive to the variations of $\Omega_\mathrm{b}$. We also note that the highest sensitivity to the variations in $\Omega_\mathrm{b}$ is not aligned with the angular velocity that matches the resonances. The sensitivity is highest up to $5\ \mathrm{km\, s^{-1}\,kpc^{-1}}$ above the resonance velocity. This is most pronounced in clusters with highly eccentric orbits. It is due to the tidal tails being more affected by the bar close to the pericentre as opposed to the apocentre or the rest of the orbit. We show that this is indeed the case in the next simulation, which explores the correlation between the shape of the tidal tails, the bar pattern speed, and the orientation of the bar. 

In simulations of clusters with small pericentres, we also observed an increase in sensitivity near the $1:4$ resonance, which has angular velocity between the CR and OLR. We further explore this resonance in simulations of real open clusters in Section \ref{sec:real_clusters}, for which the effect of the $1:4$ resonance is more pronounced than here.

In the second simulation, we kept the bar pattern speed fixed at $\Omega_\mathrm{b}=39.0\ \mathrm{km\, s^{-1}\,kpc^{-1}}$ and varied the position angle of the bar within $0.25\leq PA \leq 0.6$ (in radians). We used the same $5\times5$ grid of present-day velocities of cluster centres as in the previous simulations. The similarities between each pair of clusters are shown in Figure \ref{fig:bar_sim} (right). Both KLD and MMD distances show a mostly linear relationship between the $PA$ difference and the shapes of tidal tails, with no distinct position angles that would affect the shape of the tails more than others.

\begin{figure*}[!ht]
\begin{minipage}[t]{12cm}
    \includegraphics[height=8.05cm]{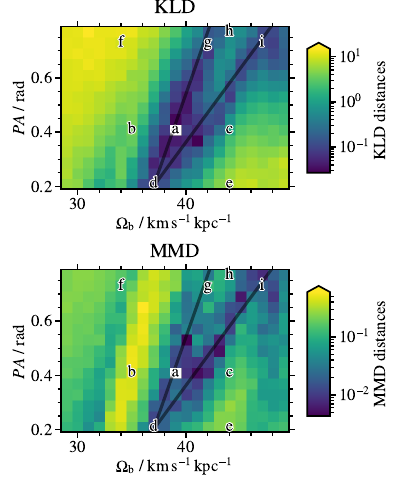} \includegraphics[height=8.05cm]{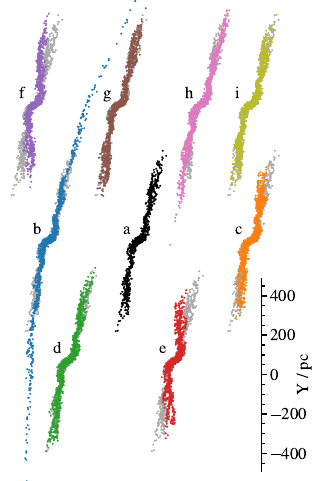}
    \end{minipage}\hfill
      \begin{minipage}[b]{6cm}
    \caption{Left: similarities between the tidal tails developed in a simulation with $v_\mathrm{R}=0.0\ \mathrm{km\,s^{-1}}$, $v_\mathrm{T}=-30.0\ \mathrm{km\,s^{-1}}$, $\Omega_\mathrm{b}=39.0\ \mathrm{km\,s^{-1}\,kpc^{-1}}$, and $PA=0.41$ and simulations with the same $v_\mathrm{R}$ and $v_\mathrm{T}$ but with $0.21\leq PA \leq0.77$ and $29.0\ \mathrm{km\,s^{-1}\,kpc^{-1}} \leq \Omega_\mathrm{b} \leq39.0\ \mathrm{km\,s^{-1}\,kpc^{-1}}$. KLD distances and MMD distances are plotted in the top and bottom panels, respectively. Black lines mark slopes of $0.114\ \mathrm{rad/km\,s^{-1}\,kpc^{-1}=111.5\ \mathrm{Myr}}$ and $0.054\ \mathrm{rad/km\,s^{-1}\,kpc^{-1}=52.8\ \mathrm{Myr}}$. Right: The structure of some simulated clusters is plotted in the Galactic $XY$ plane. The parameters of the shown clusters are marked with a -- i on the left plots. Cluster a is also plotted in grey behind every other one. All clusters are plotted at the same scale, as indicated in the bottom-right corner.}
    \label{fig:bar_cor}
    \end{minipage}
\end{figure*}

We further explored the correlation between the bar's position angle and its pattern speed. If the alignment of the bar with the position of the cluster at pericentre is critical for shaping their orbits, we expect that similar tidal tails will be produced in simulations where the $PA$ of the bar is linearly proportional to its $\Omega_\mathrm{b}$, so the cluster always meets the end of the bar close to the pericentre. A similar effect might be achieved if the bar rotates at a rate such that the opposite side of the bar is aligned with the cluster at its pericentre.  

\begin{figure}[!ht]
    \centering
    \includegraphics[width=0.99\columnwidth]{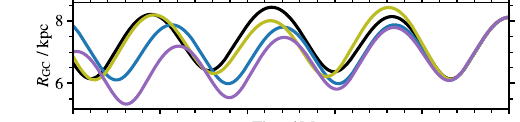}
    \includegraphics[width=0.99\columnwidth]{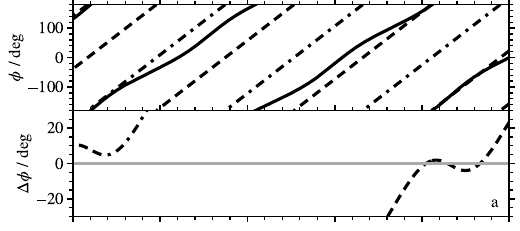}
    \includegraphics[width=0.99\columnwidth]{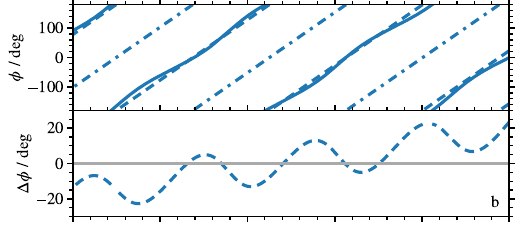}
    \includegraphics[width=0.99\columnwidth]{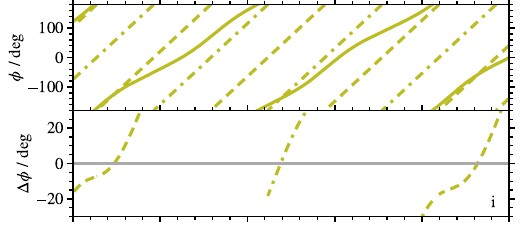}
    \includegraphics[width=0.99\columnwidth]{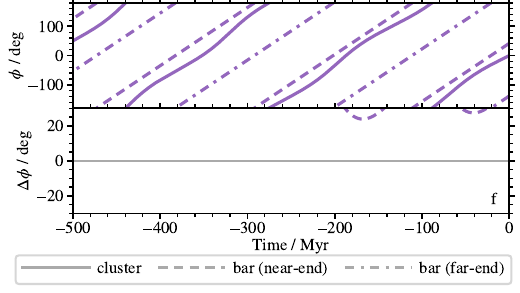}
    \caption{Orbits of clusters in four simulations marked a, b, i, and f in Figure \ref{fig:bar_cor}. Top: Galactocentric radius of cluster centres vs. time in all four simulations. Following top to bottom: Cylindrical coordinate $\phi$ of the cluster centre (solid line), near-end of the bar (dashed line), and far-end of the bar (dash-dotted line). A full range of angles is plotted, as well as the difference between the angle of the cluster centre and the ends of the bar. Panels show the case for simulations a (black), b (blue), i (yellow), and f (purple). We used the same colours as in Figure \ref{fig:bar_cor}.}
    \label{fig:bar_cor_examples}
\end{figure}

To demonstrate that $\Omega_\mathrm{b}$ and $PA$ are correlated, we plot in Figure \ref{fig:bar_cor} a similarity between the tidal tails from a reference simulation and a grid of simulations with varying $29\ \mathrm{km\, s^{-1}\,kpc^{-1}}\leq\Omega_\mathrm{b}\leq49\ \mathrm{km\, s^{-1}\,kpc^{-1}}$ and $0.2\leq PA\leq0.75$. The reference against which we computed all similarities has $\Omega_\mathrm{b}=39\ \mathrm{km\, s^{-1}\,kpc^{-1}}$ and $PA=0.41$. All simulations were done for a cluster with $v_\mathrm{R}=0.0\ \mathrm{km\,s^{-1}}$, $v_\mathrm{T}=-30.0\ \mathrm{km\,s^{-1}}$. We find that the correlation is not trivial with two obvious linear relations reflected in the grid of similarities: $\Delta PA / \Delta \Omega_\mathrm{b}=0.114\ \mathrm{rad/km\,s^{-1}\,kpc^{-1}=111.5\ \mathrm{Myr}}$ and $\Delta PA / \Delta \Omega_\mathrm{b}=0.054\ \mathrm{rad/km\,s^{-1}\,kpc^{-1}=52.8\ \mathrm{Myr}}$. The slopes, expressed in units of time, indicate the ages at which the bar and cluster were last aligned in the pericentre. We show in Figure \ref{fig:bar_cor_examples} that this has indeed happened around the times indicated by the slopes.

The degeneration between the pattern speed and the position angle of the bar can be resolved if multiple clusters are observed. The correlation in Figure \ref{fig:bar_cor} shows that it is important that the bar catches the cluster in its pericentre at the correct time. Clusters with different angular speeds and different positions relative to the bar's $PA$ cannot all reach the pericentre at the same time or in the same orientation relative to the bar. Observing only a small number of well-selected clusters would hence break the degeneracy. In other words, a non-linear effect of the bar pattern speed on the shape of the tidal tails in Figure \ref{fig:bar_sim} (left) and a linear effect of the bar's position angle in Figure \ref{fig:bar_sim} (right) can be separated. 

\begin{figure}[!ht]
    \centering
    \includegraphics[width=\columnwidth]{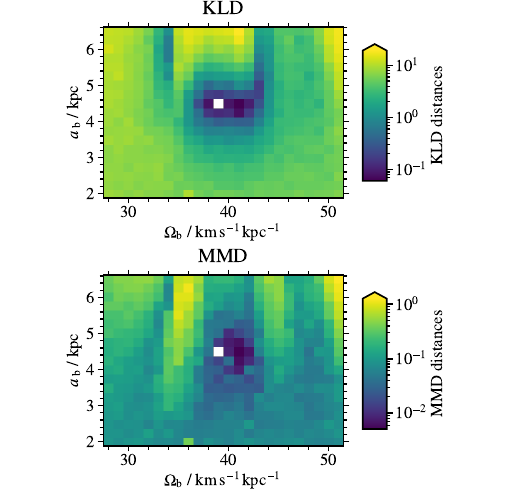}
    \caption{Similarities between the tidal tails developed in a simulation with $v_\mathrm{R}=0.0\ \mathrm{km\,s^{-1}}$, $v_\mathrm{T}=-30.0\ \mathrm{km\,s^{-1}}$, $\Omega_\mathrm{b}=39.0\ \mathrm{km\,s^{-1}\,kpc^{-1}}$, and bar's semi-major axis $a_\mathrm{b}=4.5\ \mathrm{kpc}$ and simulations with $2.0\ \mathrm{kpc}\leq a_\mathrm{b} \leq6.5\ \mathrm{kpc}$ and $28.0\ \mathrm{km\,s^{-1}\,kpc^{-1}} \leq \Omega_\mathrm{b} \leq51.0\ \mathrm{km\,s^{-1}\,kpc^{-1}}$. KLD distances and MMD distances are plotted in the top and bottom panels, respectively.}
    \label{fig:bar_cor_2}
\end{figure}

Literature shows a dichotomy or an anti-correlation between the pattern speed of the bar and the length of the bar; the bar is short and fast or long and slow. To explore this anti-correlation in our simulations, we plotted the similarities between the tidal tails in a reference simulation and a grid of simulations with varying $\Omega_\mathrm{b}$ and length of the bar $a_\mathrm{b}$ (Figure \ref{fig:bar_cor_2}). All simulations were done for the same $v_\mathrm{R}=0.0\ \mathrm{km\,s^{-1}}$, $v_\mathrm{T}=-30.0\ \mathrm{km\,s^{-1}}$, and $PA=0.41$. The reference against which the simulations in a grid were compared had $\Omega_\mathrm{b}=39.0\ \mathrm{km\,s^{-1}\,kpc^{-1}}$, $a_\mathrm{b}=4.5\ \mathrm{kpc}$. In the literature, bars with pattern speed of $\sim 35\ \mathrm{km\,s^{-1}\,kpc^{-1}}$ have lengths of $\sim5\ \mathrm{kpc}$ and bars with pattern speed $\sim 55\ \mathrm{km\,s^{-1}\,kpc^{-1}}$ have lengths of $\sim3\ \mathrm{kpc}$. We observe no such anti-correlation in our simulations. In the particular case shown in Figure \ref{fig:bar_cor_2}, the tidal tails are indistinguishable from the reference tidal tails in a region $\Delta \Omega_\mathrm{b}=3\ \mathrm{km\,s^{-1}\,kpc^{-1}}$ and $\Delta a_\mathrm{b}=0.5\ \mathrm{kpc}$ and outside these limits, we observe no anti-correlation similar to the one implied in the literature. 

\subsection{Giant molecular clouds}

We performed simulations with the same $5\times5$ grid of present-day $v_\mathrm{R}$ and $v_\mathrm{T}$. For each velocity, we simulated one cluster with ten randomised positions for the GMCs. Each cluster initially included 4000 stars. The total mass of the GMCs was $3.1\,10^{8}\ M_\odot$ \citep{rice16}. The simulation was run for $500\ \mathrm{Myr}$. Because the simulated clusters orbit in the Galactic plane, where the density of GMCs is the largest, we expect that the perturbations are more frequent than for a real population of open clusters. On the other hand, many open clusters are older than $500\ \mathrm{Myr}$, so the perturbations of GMCs on those clusters are more numerous. Nevertheless, here we analyse only the simulation described above. 

\begin{figure*}[!ht]
    \includegraphics[width=12cm]{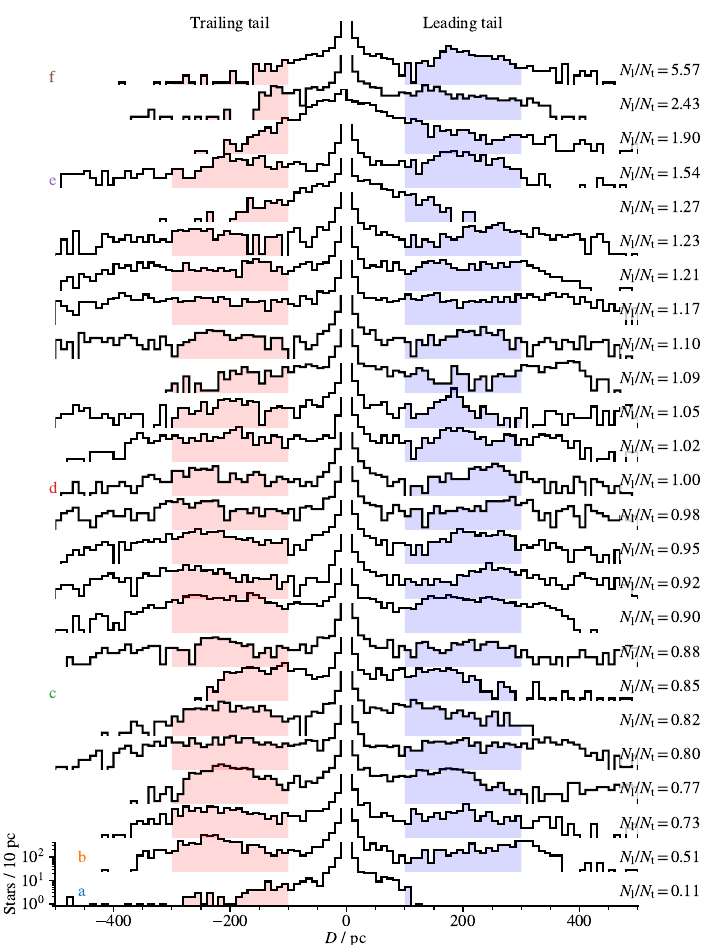}\hfill\includegraphics[width=6cm]{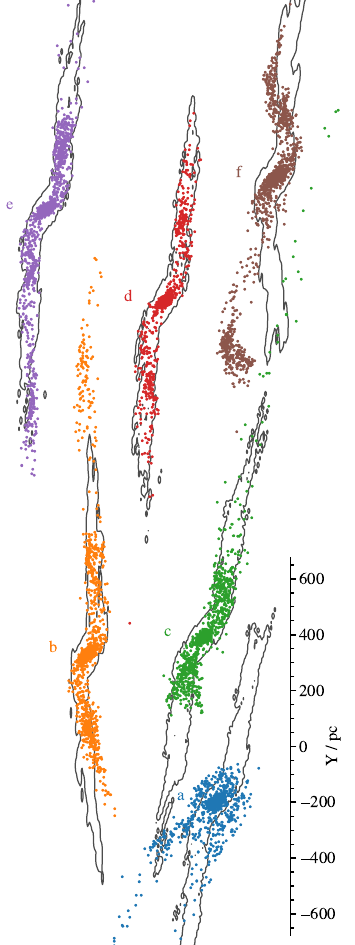}
    \caption{Left: Stellar densities along the principal axis of 25 simulated clusters. Plots for individual clusters are shifted vertically for clarity and share the same scale as shown for the bottom-most plot. $D$ is the position along the principal axis of the cluster. Right: six clusters marked a to f in the left plot are shown in the XY plane. They all share the same scale marked on the bottom-right. Images of clusters are intentionally scattered to avoid overlaps. The underlying contours show the outline of the tidal tails if there was no perturbation by GMCs.}
    \label{fig:asym}
\end{figure*}

In a visual inspection of our simulations, the clusters and tidal tails look remarkably similar to the real-life tails recovered in \citet{kos24}. In \citet{kos24}, we showed that it is possible to find stars in the tidal tails far from the cores of open clusters, provided we assume a cluster dissolution model, the Galactic gravitational potential, and a population model of the Milky Way. None of these models includes physics to explain asymmetric tidal tails; however, when applied to Gaia data, we observe that most clusters show a detectable asymmetry in the number of stars between the leading and trailing tails. In \citet{kos24}, we concluded that observed asymmetries and the diverse structure of tidal tails prove that the data (as opposed to the models/priors) are driving the shape of the tidal tails and the membership probabilities in open clusters.

The effect of GMCs in our simulations is illustrated in Figure \ref{fig:asym}. On the left-hand side we show stellar densities along the principal axis of 25 simulated clusters. These can be directly compared with plots in \citet{kos24}, such as the one in the second panel of Figure A.1 in \citet{kos24}. On the right of each panel, we list the asymmetry index (number of stars in the leading tail divided by number of stars in the trailing tail, both measured within the regions shaded in red and blue in Figure \ref{fig:asym}). On the right-hand side we show six clusters in the XY plane, all at the same scale. Cluster d appears unperturbed by the GMCs, and the other five show different degrees of perturbations, either producing tails with a heavy trailing tail (b, c) or a heavy leading tail (e, f). A chaotic system of tails is shown for cluster a.

\subsection{Satellite galaxies}

\begin{figure}[!ht]
        \includegraphics[width=\columnwidth]{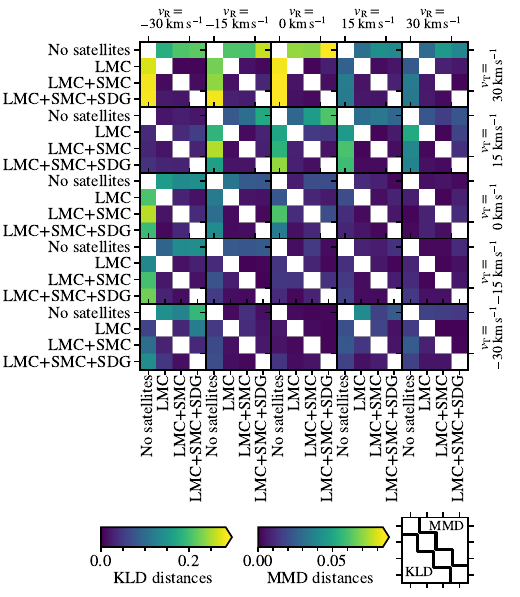}
        \caption{Similarities between all pairs of simulations with no satellite galaxies, with LMC, SMC, and SDG. 25 panels show a combination of present day velocities $v_\mathrm{R}$ (left to right) and $v_\mathrm{T}$ (top to bottom). Above the diagonals in each panel, we plot the MMD distance; below the diagonals, the KLD distance. We note that the numerical values for MMD and KLD distances are not on the same scale, hence two colour bars.}
    \label{fig:satellites_sim}
\end{figure}

We explored the effect of satellite galaxies LMC, SMC, and SDG in a simulation with a duration of $1.5\ \mathrm{Gyr}$. We detected a small difference between the clusters evolved in a potential with no satellites and in a potential including the LMC. No differences were detected when adding the other two satellite galaxies. Similarities between the simulations are illustrated in Figure \ref{fig:satellites_sim}. The differences are most prominent for clusters on orbits with a large apocenter. Although the differences are detectable with statistical metrics, a visual inspection indicates that satellite galaxies have a negligible effect on the evolution of tidal tails in open clusters. We assume that such small differences would be undetectable in real data. Possible exceptions would be clusters on particularly extreme orbits that bring them close to the satellite galaxies, which we do not explore in this work.

\section{Simulations of real open clusters}
\label{sec:real_clusters}

In the previous sections, we analysed completely synthetic clusters. They had orbits aligned with the Galactic plane, and all were positioned at Solar galactic coordinates at the present time. Synthetic clusters were realistic in terms of cluster sizes and had ages and orbital eccentricities that are typical for real open clusters. In this section, we extend our analysis to the tidal tails of real open clusters catalogued in the Solar neighbourhood. In Section \ref{sec:sensitivity}, we have shown that the pattern speed of the bar has a significant effect on the shapes of the tidal tails, so in this section, we analyse only the sensitivity of real open clusters on the pattern speed. 

\subsection{Data}

To select clusters suitable for our analysis, we used the catalogue of open clusters produced by \citep{hunt24}. We selected clusters located within 1 kpc from the Sun. We further limited the selection to clusters with age between $250\ \mathrm{Myr}$ and $3\ \mathrm{Gyr}$. Clusters develop the tidal tails gradually, and we expect the clusters younger than $250\ \mathrm{Myr}$ to have such short tidal tails that they could not have been perturbed by the bar yet. Clusters older than $3\ \mathrm{Gyr}$ are hard to simulate, because in our simulations they completely dissolve. This can be mitigated by initialising a very large cluster with $>5000$ stars, which slows down the simulation considerably. A list of selected clusters is given in Appendix \ref{sec:clusters_table_appendix} (Table \ref{tab:clusters}). In the table, we provide the boundary conditions (current positions and velocities), from which we computed the initial conditions for the simulation and some basic orbital parameters (pericentre, apocentre distances, maximum distance from the Galactic plane, orbital eccentricity, and mean angular velocity). We computed the initial conditions the same way as for synthetic clusters by integrating the orbit of the cluster centre back in time to the cluster's age. We then constructed a cluster at that position and evolved it with an n-body simulation. We initialised the clusters with different numbers of stars, based on the cluster's age. For clusters younger than $1.5\ \mathrm{Gyr}$ we used 2000 stars, while for older clusters we used 4000 stars. We kept other parameters (radius, IMF, virial ratio) the same as in the simulations of synthetic clusters.

We only performed simulations with one set of boundary conditions for each cluster (RA, declination, proper motions in both directions, distance, radial velocity and age). We did not propagate their uncertainty into our simulations. Radial velocity is occasionally measured with poor accuracy in \citet{hunt24}; either the number of radial velocity measurements for cluster stars is too low to provide a reliable mean value, or radial velocities of individual stars are distorted by binaries. Of the total of 161 clusters in our sample, 28 either have a standard deviation of radial velocity greater than $20\ \mathrm{km\,s^{-1}}$ or the number of stars used to calculate their radial velocity is less than ten. The age of the clusters has by far the largest relative uncertainty among the parameters, often in the order of 50\%. We used the mean value (\texttt{logAge50}). In the second paper, we will explore the impact of uncertainties on the tidal tails and their observability.

\subsection{Analysis of tidal tails of real open clusters}

We simulated the development of tidal tails in the common gravitational potential where we varied the bar pattern speed between $\Omega_\mathrm{b}=28.0\ \mathrm{km\,s^{-1}\,kpc^{-1}}$ and $\Omega_\mathrm{b}=51.0\ \mathrm{km\,s^{-1}\,kpc^{-1}}$ in steps of $1\ \mathrm{km\,s^{-1}\,kpc^{-1}}$, same as in the exploration of synthetic clusters. We show the simulated tidal tails of ten clusters with significant dependence on the pattern speed in Figure \ref{fig:real_clusters}. All simulated clusters are shown in Appendix \ref{sec:additional_figures_appendix}. We plot the shape of the tidal tails only in the $XY$ plane, because we found that the tails simulated with different $\Omega_\mathrm{b}$ have no significant spread in the $Z$ direction.

\begin{figure*}[!p]
    \includegraphics[scale=1]{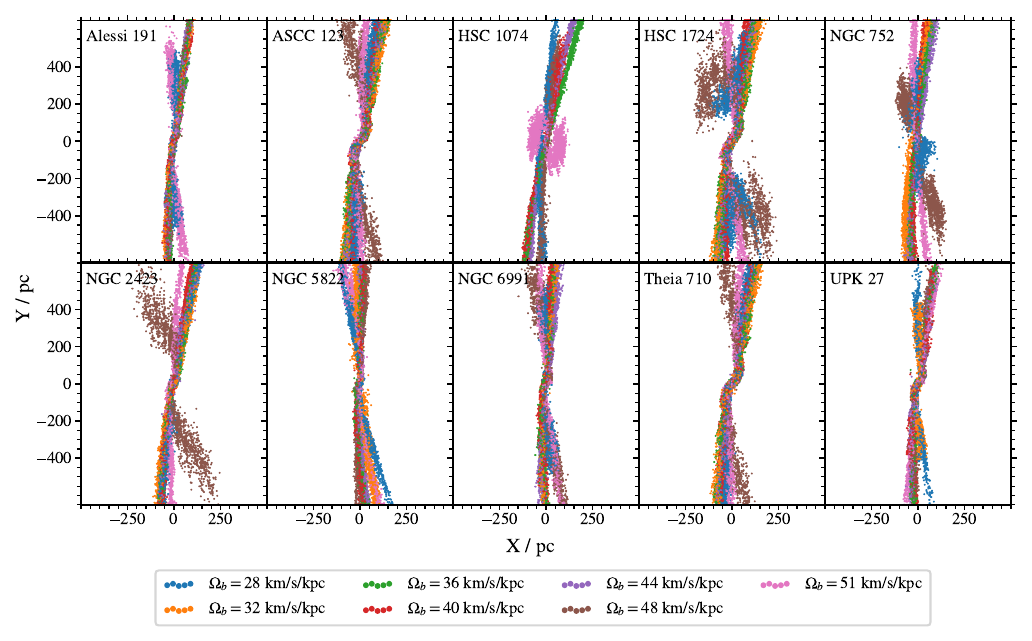}
    \caption{Simulations of tidal tails for selected real open clusters. In each panel, we plot a cluster's tidal tails in the Galactic plane, for 7 different values of bar pattern speed $\Omega_\mathrm{b}$. Stars in each panel are plotted in random order. Simulation results for all simulated clusters can be found in Appendix \ref{sec:additional_figures_appendix}.}
    \label{fig:real_clusters}
\end{figure*}

\begin{figure*}[!p]
    \includegraphics[scale=1]{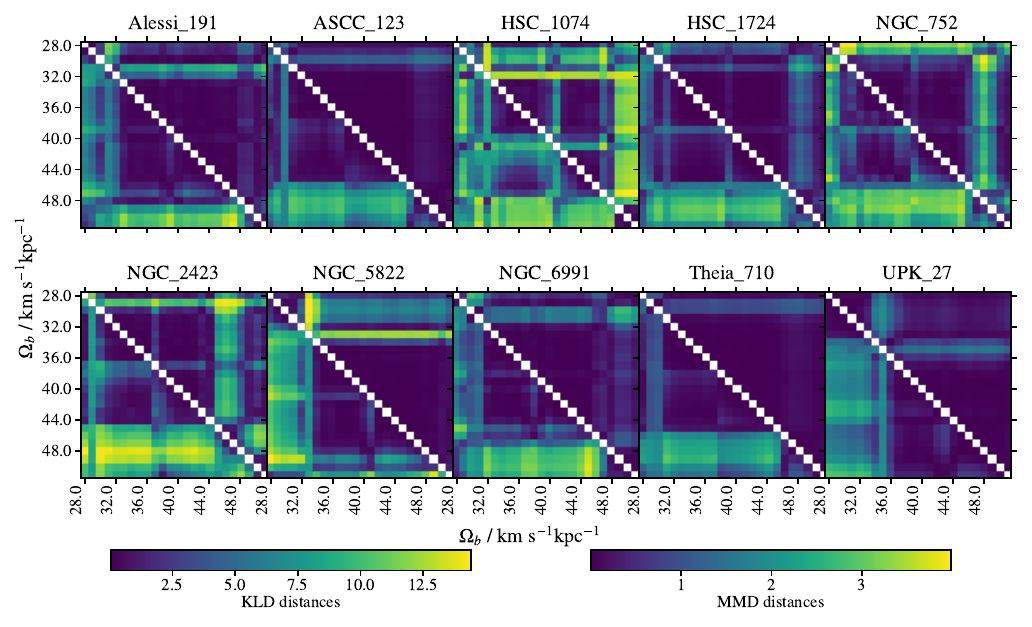}
    \caption{Similarities between all pairs of simulations with varied bar pattern speed $\Omega_\mathrm{b}$ for clusters shown in Figure \ref{fig:real_clusters}. In each panel, we plot KLD (below the diagonal) and MMD (above the diagonal) distances for each pair of simulations in a selected cluster. We note that the numerical values for the MMD and KLD distances are not on the same scale; hence, two colour bars.}
    \label{fig:real_clusters_similarities}
\end{figure*}

Same as for synthetic clusters, we observe that tidal tails of real open clusters change orientation of the tails and their length when we vary the bar pattern speed. Close to the resonance pattern speeds, they also exhibit folding of the tails (e.g., in HSC 1074, HSC 1724, and NGC 752, some showing folding at multiple resonances), as we have observed near the resonances in previous simulations. Figure \ref{fig:real_clusters_similarities} illustrates the similarities between tidal tails developed in simulations with a varying $\Omega_\mathrm{b}$. Panels show the grids for the same ten clusters displayed in Figure \ref{fig:real_clusters}. 

We notice that the shapes of the tidal tails are most sensitive to the variations of $\Omega_\mathrm{b}$ close to the CR. The studied clusters are typically in CR with the bar, if it has a pattern speed between $25\ \mathrm{km\,s^{-1}\,kpc^{-1}} \leq\Omega_\mathrm{b}\leq 35\ \mathrm{km\,s^{-1}\,kpc^{-1}}$. Clusters can be in a Lindblad resonance \citep{binney08} when the bar pattern speed equals
\begin{equation}
    \Omega_b = \Omega_{\phi} \pm \frac{1}{2}\Omega_R,
\end{equation}
where $\Omega_{\phi}$ and $\Omega_R$ are the angular velocity and radial frequency of a cluster's orbit. A positive sign means an OLR. We do not see the effect of the inner Lindblad resonance because, in the Solar neighbourhood, it would have occurred at unrealistic pattern speeds not simulated here. For clusters studied in this work, the OLR happens when $40\ \mathrm{km\,s^{-1}\,kpc^{-1}} \leq\Omega_\mathrm{b}\leq 55\ \mathrm{km\,s^{-1}\,kpc^{-1}}$. 

In Figure \ref{fig:real_clusters_similarities}, we can clearly see that tidal tails are sensitive to both the CR and OLR. We also notice some sensitivity to the pattern speeds between the CR and the OLR, which we computed to correspond to the $1:4$ resonance where 
\begin{equation}
    \Omega_b = \Omega_{\phi} \pm \frac{1}{4}\Omega_R.
\end{equation}
In Figure \ref{fig:real_clusters}, we can attribute every tidal tail that stands out from the rest to the cluster on an orbit close to one of the CR, OLR or the $1:4$ resonance. Hence, the high sensitivity of the tidal tails to varying pattern speeds near the resonances is not detected just numerically, but the signature is visually significant as well. Orientation of the tails and folding (when the pattern speed is very near the resonant speed) make the biggest visual difference; however, the shape of the tails and structures and overdensities inside the tails vary as well. These are harder to distinguish visually, but are picked up by the distance metrics. The lengths of the tails (cropped in Figure \ref{fig:real_clusters}) vary as well and contribute to the distances computed by KLD and MMD. We note that in observations of real clusters, the lengths of the tails would be hardest to measure because the lengths of gradually less dense tails are poorly defined, and stars far from the cluster cores are nearly impossible to find. Luckily, in simulations where the length of the tails changes, so do the orientation and shape.

\begin{figure*}[!ht]
    \includegraphics[scale=1]{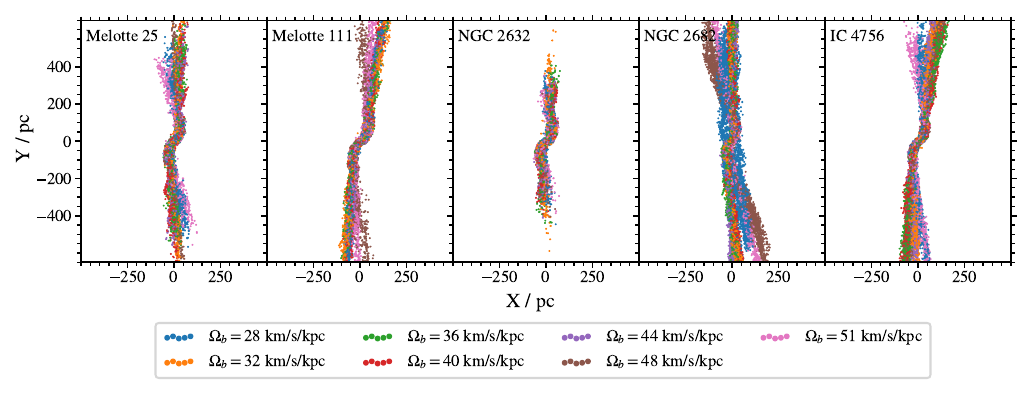}
    \caption{Simulation results for clusters whose tidal tails have been studied in the literature. In each panel, we plot a cluster's tidal tails in the Galactic plane, for 7 different values of bar pattern speed $\Omega_b$. Stars in each panel are plotted in random order. Simulation results for all simulated clusters can be found in Appendix \ref{sec:additional_figures_appendix} (see Figures \ref{fig:real_clusters_a}--\ref{fig:real_clusters_e}).}
    \label{fig:clusters_lit}
\end{figure*}

Tidal tails of several clusters simulated in this work have been studied observationally in the literature. Here we discuss the shape of tidal tails and their sensitivity to the bar pattern speed for six such clusters. In Figures \ref{fig:real_clusters} and \ref{fig:real_clusters_similarities} we summarise the results for NGC~752, and in Figure \ref{fig:clusters_lit}, we show the simulated shapes of the tails for five more. One of the most researched is the Hyades cluster (Melotte 25). Our simulations show extended tidal tails, which vary in length and orientation and are sensitive to many distinct values of $\Omega_\mathrm{b}$. The sensitivity of Hyades' tidal tails to the bar has been studied by \citet{zhou26}. Multiple papers report asymmetry in the Hyades' tidal tails \citep[e.g.][]{jerabkova21, kroupa22}. We see no obvious asymmetry in our simulations, from which we conclude that the bar (which can induce small asymmetry) is not responsible for such effects. Other clusters with observed tail asymmetry include Melotte 111, NGC 2632 and NGC 752 \citep{kroupa22, boffin22}, only the latter showing asymmetric tidal tails in our simulations. Simulations of Melotte 111 show that the orientation of the tails has a small sensitivity on $\Omega_\mathrm{b}$ close to its OLR. NGC 2632 shows almost no sensitivity on $\Omega_\mathrm{b}$, and in NGC 752 we detect strong sensitivity around CR and OLR, specifically for $\Omega_b=28\,\mathrm{km\,s^{-1}\,kpc^{-1}}$ and  $\Omega_b=48\, \mathrm{km\,s^{-1}\,kpc^{-1}}$, as well as small sensitivity at $1:4$ resonance. We found asymmetric and distorted tails, with a longer leading tail for $\Omega_b=28\,\mathrm{km\,s^{-1}\,kpc^{-1}}$ and a longer trailing tail for $\Omega_b=48\, \mathrm{km\,s^{-1}\,kpc^{-1}}$. Clusters IC 4756 with tidal tails recovered in \citet{ye21} and NGC 2682 (M67) with tidal tails recovered in \citet{gao20_pasj} show extended tidal tails with strong sensitivity to $\Omega_\mathrm{b}$.

\section{Conclusions and discussions}
\label{sec:conclusions}


The shape of tidal tails of open clusters in the Solar neighbourhood depends greatly on the gravitational potential of the Milky Way. We built the gravitational potential with a combination of simple potentials with a few free parameters. By varying each parameter of the potential independently, we were able to systematically study their effects on the tidal tails. We found that the bar dominates over the effects of spirals and satellite galaxies and competes with the GMCs. In this work, we aimed to study the sensitivity of the tidal tails to variations in the bar pattern speed, which is the most intriguing parameter that could be measured from the shape of the tidal tails of open clusters. From computed statistical distances between simulated tidal tails and after visual inspection of the shape of the tails, we conclude that differences in pattern speed in the order of $1\ \mathrm{km\, s^{-1}\, kpc^{-1}}$ can be observed.

The primary signature of the bar pattern speed in open clusters is the orientation of the tidal tails. The bar affects the orientation and eccentricity of the cluster's orbit. The orientation of the tails is aligned with the orientation of the orbit so that we can detect differences between orbits just by the orientation of the tails, regardless of the length or structure in the tails. Orientation of the orbit is also little affected by weak interactions with GMCs, which might scatter some stars and disrupt the structures in the tails, but cannot change the energy or angular momentum of the cluster's orbit as much as the bar does. Strong interactions where the orbit of the cluster (and the stars already in the tidal tails) gets perturbed can only happen by the most massive GMCs in orbits that are similar to the orbit of the cluster. This is unlikely to happen to most clusters. We observe that even clusters that experienced weak perturbations by GMCs conserve the orientation of the tidal tails. 

\citet{moreira25} found that objects in orbits similar to the Sun encounter a GMC once every $700\ \mathrm{Myr}$. Most of our simulated clusters experienced at least one encounter in $500\ \mathrm{Myr}$. The clusters with a system of tidal tails hundreds of parsecs long are much larger structures, of course, hence they are perturbed by more GMCs. As we qualitatively found out, the GMCs rarely perturb the whole system of tails, so with careful consideration of possible perturbation by a GMC, the observed tidal tails could reliably be used to infer the pattern speed of the bar. We must, however, be cautious when studying the clusters systematically and not draw conclusions based on extreme cases such as the Hyades, with their highly asymmetric tidal tails. 

The internal structure of the tails across different simulations can be quantitatively compared using statistical metrics, of which we had the most success with KLD and MMD. Position of the epicyclic overdensities is dictated by the shape of the orbit and position of the cluster on the orbit \citep{kupper08}, hence the signature of the bar is reflected in the structure of the tails as well. The internal structure will be much harder to use when inferring the bar pattern speed from observations. We can produce simulations where stars densely populate tidal tails; however, on the sky, observational limitations, biases, and uncertainties might prevent us from finding as many stars as we used in our simulations (typically, we simulated 2000 stars, including those in the cluster core).

\begin{figure}[!ht]
    \centering
    \includegraphics[width=0.99\columnwidth]{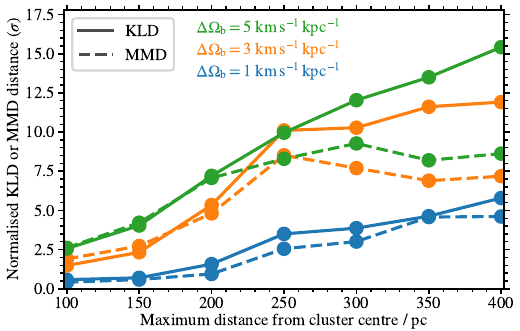}
    \caption{Sensitivity of tidal tails to the variations of the bar pattern speed when stars far from the cluster are discarded. The plot shows the case for a synthetic cluster in a CR with the bar at $v_\mathrm{R}=0\ \mathrm{km\, s^{-1}}$, $v_\mathrm{T}=-30\ \mathrm{km\, s^{-1}}$, and $\Omega_\mathrm{b}=34.0\ \mathrm{km\, s^{-1}\, kpc^{-1}}$. To compute the similarities between this cluster and a cluster where $\Omega_\mathrm{b}$ is 1, 3, or 5 $\mathrm{km\, s^{-1}\, kpc^{-1}}$ different, we only used stars up to the maximum distance from the cluster core. Similarities are given in standard deviations. Solid lines show results for the KLD metric and dashed lines for the MMD metric.} 
    \label{fig:bar_max}
\end{figure}

In real data, we do not expect the full extent of the tidal tails to be observed. Stars closer to the core are more likely to be recovered reliably than those far away. If varying parameters of the bar only affected the far ends of the tails, our endeavour would not have real applications. For this discussion, we computed similarities only from stars near the cluster core and checked whether they still show sensitivity to a varying bar pattern speed. We chose a synthetic cluster with $v_\mathrm{R}=0\ \mathrm{km\, s^{-1}}$, $v_\mathrm{T}=-30\ \mathrm{km\, s^{-1}}$, and used as a reference a simulation where $\Omega_\mathrm{b}=34.0\ \mathrm{km\, s^{-1}\, kpc^{-1}}$, so the cluster is close to the CR of the bar. Then we computed similarities with simulations of clusters where $\Omega_\mathrm{b}$ is one, three or five $\mathrm{km\, s^{-1}\, kpc^{-1}}$ larger or smaller. Each time we used only stars within a certain distance from the cluster to compute the similarities, up to $400\ \mathrm{pc}$, which is approximately the length of each tail. Figure \ref{fig:bar_max} provides the results. We averaged two similarities for $\Omega_\mathrm{b}$ symmetrically larger or smaller than the reference. We computed both KLD and MMD distances and expressed them in units of standard deviations. 

In Figure \ref{fig:bar_max} we see that differences in $\Omega_\mathrm{b}$ of a few $\mathrm{km\, s^{-1}\, kpc^{-1}}$ are significant, even if we only take stars $d<200\ \mathrm{pc}$ from the core into the account. To detect a difference of only $1\ \mathrm{km\, s^{-1}\, kpc^{-1}}$, we should observe the whole extent of the tidal tails. Our findings are in agreement with \citet{zhou26}, where they find that the bar pattern speed can be roughly measured from stars in Hyades that are closer than $170\ \mathrm{pc}$ from the core, assuming the cluster is also close to the CR. Our results in Figure \ref{fig:bar_max} are also plotted for a cluster close to the resonance. It is obvious from Figure \ref{fig:bar_sim} that, looking at a cluster far from the resonance, the conclusions would be much less optimistic. This is expected, as we noticed that clusters far from resonances have tidal tails that are completely insensitive to variations in the bar pattern speed, regardless of how far from the cluster we observe the tails. 

\begin{figure}[!ht]
    \centering
    \includegraphics[width=0.99\columnwidth]{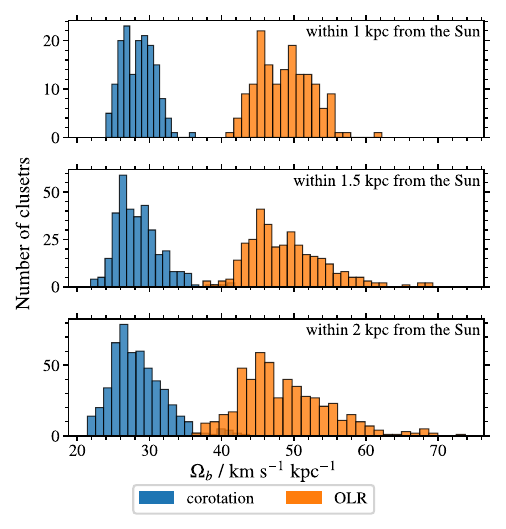}
    \caption{Distribution of CR and OLR for open clusters in the Solar neighbourhood, assuming a bar pattern speed given on the horizontal axis.}
    \label{fig:available_clusters}
\end{figure}

In this work we explored clusters with tidal tails that are sensitive to the gravitational effects of the bar with the aim of once using the tails to measure the pattern speed of the bar. We showed that clusters close to the resonances are most suitable. We also showed that tidal tails are sensitive to the resonances even when the cluster's mean angular velocity is up to $5\ \mathrm{km\,s^{-1}\,kpc^{-1}}$ from the resonant velocity. Because a typical angular velocity in the Solar neighbourhood is around $28\ \mathrm{km\,s^{-1}\,kpc^{-1}}$, the tidal tails of nearby clusters are sensitive to variations of the bar pattern speed that is close to $\sim28\ \mathrm{km\,s^{-1}\,kpc^{-1}}$ or $\sim50\ \mathrm{km\,s^{-1}\,kpc^{-1}}$, corresponding to the CR and OLR, respectively. This leaves a gap of pattern speeds at around $40\ \mathrm{km\,s^{-1}\,kpc^{-1}}$, which we can not investigate via open clusters in the Solar neighbourhood, because no cluster has a resonant angular velocity close to this value. The distribution of resonant angular velocities for clusters within $1.0\ \mathrm{kpc}$,  $1.5\ \mathrm{kpc}$, and $2.0\ \mathrm{kpc}$ from the Sun is shown in Figure \ref{fig:available_clusters}. We will need to observe clusters up to $2.0\ \mathrm{kpc}$, if we want to exploit the CR and OLR to measure the bar pattern speed. 

In the gap between the CR and OLR, we can also exploit the $1:4$ resonance. We observed in many clusters that the tidal tails are sensitive to small variations of the bar pattern speed around this resonance. Unlike for the CR and OLR, the sensitivity is limited to around $\pm 1\ \mathrm{km\,s^{-1}\,kpc^{-1}}$ around the $1:4$ resonance, and the effect of the bar appears to be smaller than for clusters in the CR or OLR. Nevertheless, the $1:4$ resonance covers the gap between the CR and OLR even for clusters within $1.0\ \mathrm{kpc}$ from the Sun. We conclude that there are enough nearby clusters and their orbits are diverse enough that the bar pattern speed could be constrained by observations within $1.0\ \mathrm{kpc}$ from the Sun.  

While we focus on the bar pattern speed in this work, the tidal tails of clusters are sensitive to other parameters of the bar. Size, structure described by different modes \citep{khalil25}, possibly each with a different pattern speed, and bar deceleration \citep{chiba21b}. The influence of all these on the tidal tails of open clusters could be studied in simulations like ours; however, before the current global bar pattern speed is actually observed via the shape of the tidal tails, the observability of any other complication is rather speculative. 


We show that at least half of the real open clusters studied in this work are inherently unaffected by the variations of the bar pattern speed. A similar fraction was also identified by \citep{jadhav25}. Such clusters should not be dismissed as insignificant. Because of the trivial nature of their tidal tails, we can find stars in the tails of such clusters by modelling their dissolution and searching for stars in the parameter space indicated by a simulation as presented in \citet{kos24}. A simulation that assumes an incorrect gravitational potential will still yield reliable tidal tails and help us find stars far from the cluster cores. 


In this work, we computed all similarities between tidal tails in a 2D projection onto the Galactic plane. While the algorithms we used work well in any number of dimensions, we did not find any improvement when similarities were computed in 3D, even for clusters that orbit well outside the Galactic plane. It appears that all the information that defines the difference between the shapes of tidal tails of open clusters is contained in the X-Y positions of stars. Because the tidal tails are long, thin structures, their positions in Z are highly correlated with those in the X-Y plane. Thus, they cannot contribute any additional fidelity to the similarity metrics. On the other hand, velocities could be included, which we will do in the second paper. In particular, the observed velocities (proper motions and radial velocities) can exhibit complex patterns along the tails due to a combination of projection effects and galactic dynamics \citep{jerabkova21, kos24, zhou26}. Hence, the velocities, used together with positions, make a parameter space in which the similarities should be computed.

\subsection*{Data availability}
Code to reproduce our Galactic gravitational potential is given in Appendix \ref{sec:potential_appendix}. Simulations of synthetic clusters will be shared via private communication upon a reasonable request.

\begin{acknowledgements}
JK and SI are supported by the Slovenian Research Agency ARIS grant P1-0188.  This work has made use of data from the European Space Agency (ESA) mission Gaia (\url{https://www.cosmos.esa.int/gaia}), processed by the Gaia Data Processing and Analysis Consortium (DPAC, \url{https://www.cosmos.esa.int/web/gaia/dpac/consortium}). Funding for the DPAC has been provided by national institutions, in particular the institutions participating in the Gaia Multilateral Agreement.
\end{acknowledgements}

\bibliographystyle{aa}
\bibliography{bib}

\begin{appendix}
\onecolumn

\section{Gravitational potentials of the Milky Way}
\label{sec:potential_appendix}

We modelled the common gravitational potential of the Milky Way with five components: the nucleus, the bar, the thin and thick stellar disks, and the dark matter halo. The complete gravitational potential is the sum of all five components.

We represented the stellar disks by two Miyamoto-Nagai potentials. The total mass of the disk is around $5 \ 10^{10}\ M_\odot$ \cite{Licquia15}. Based on \cite{khanna25}, the thick disk has double the mass of the thin disk. The estimated radial scale of the thin disk is $3.6-4.6$ kpc and its scale height $0.18-0.3$ kpc, while the estimates for the thick disk are $2-3$ kpc for its radial scale and $0.4-0.7$ kpc for its scale height \citep{khanna25, everall22a, everall22b, bland16}. We set the mass of the thin disk to $M=1.45\,10^{10}\ M_\odot$ and the mass of the thick disk to $M=2.9\,10^{10}\ M_\odot$. The scales of the disks were set according to \citet{khanna25}.

To describe the potential of the bar, we used the softened-needle bar potential \citep{long92}. As with other components, the parameters of the bar vary significantly in the literature. \citep[e.g.][]{shen20, lucey23, bland16, portail17}. We set the mass of the bar to $M=0.6\,10^{10}\ M_\odot$, and the dimensions (half-lengths) to $a=4.5\ \mathrm{kpc}$, $b=0.9\ \mathrm{kpc}$, and $c=2.0\ \mathrm{kpc}$. We set the pattern speed in the common potential to $\Omega_\mathrm{b}=39\ \mathrm{km\,s^{-1}\,kpc^{-1}}$ and the orientation of the bar to $PA=0.41\ \mathrm{rad}=23.49^\circ$.

The dark matter halo is described using the Navarro-Frenk-White profile. The estimated scale radius of the Milky Way's dark halo is in the order of $10$ kpc \citep[e.g][]{sofue12, Lin19} with large variations in the literature. In this work, we set the scale radius of the halo to $R=12.0\ \mathrm{kpc}$. Virial radius is $R_\mathrm{vir}=320\ \mathrm{kpc}$.

To model the nucleus, we used a Plummer potential with $R=300\ \mathrm{pc}$ and set its mass such that the rotational curve in the Galactic nucleus matches the observations.

Here, we provide Python code that reproduces the common gravitational potential of the Milky Way as used in this work. Code shown in Figure \ref{fig:potential_code_1} produces the potential in physical units. Amplitudes and scales are set such that $v_\mathrm{circ}=233.4\ \mathrm{km\, s^{-1}}$ at $R_\mathrm{GC}=8.122\ \mathrm{kpc}$. Code in Figure \ref{fig:potential_code_2} gives the potential in natural units, such that  $v_\mathrm{circ}=1$ at $R_\mathrm{GC}=1$. We converted the amplitudes of potentials given in natural units using a formula
\begin{equation}
    \mathrm{amp}=\frac{GM}{R_0 V_0^2},
\end{equation}
where $M$ is the amplitude of the potential in physical units, and $R_0=8.122\ \mathrm{kpc}$ and $V_0=233.4\ \mathrm{km\, s^{-1}}$. The potential in natural units can easily be scaled by any $R_0$ and $V_0$, simply by providing the values in \textsc{Galpy} functions that have the potential as an argument. 

\begin{figure}[!ht]
\begin{lstlisting}[language=Python]
import astropy.units as u
from galpy import potential

# Potentials
thin_disk = potential.MiyamotoNagaiPotential(amp=1.45*10.**10.*u.Msun,\
                                             a=4.2*u.kpc,\
                                             b=0.18*u.kpc)

thick_disk = potential.MiyamotoNagaiPotential(amp=2.9*10.**10.*u.Msun,\
                                              a=2.7*u.kpc,\
                                              b=0.48*u.kpc)
                                              
halo = potential.NFWPotential(amp=4.757*10**11.*u.Msun,\
                              a=12.0*u.kpc)
                              
bar = potential.SoftenedNeedleBarPotential(amp=0.6*10.0**10.*u.Msun,\
                                           a=4.5*u.kpc,\
                                           b=0.9*u.kpc,\
                                           c=2.*u.kpc,\
                                           omegab=39.0*u.km/u.s/u.kpc,\
                                           pa=0.41*u.rad)
                                           
nucleus = potential.PlummerPotential(amp=1.00*10.**10*u.Msun,\
                                     b=300.*u.pc)

# Combined potential
potential = [thin_disk, thick_disk, halo, bar, nucleus]
\end{lstlisting}
\caption{Python code that reproduces the common gravitational potential of the Milky Way in physical units.}
\label{fig:potential_code_1}
\end{figure}

\begin{figure}[!ht]
\begin{lstlisting}[language=Python]
import astropy.units as u
from galpy import potential

# Example of scaling parameters.
# Distance of the Sun from the Galactic centre
RO = 8.122 * u.kpc
# Circular velocity at Sun's distance
VO = 233.4 * u.km/u.s

# Potentials
thin_disk = potential.MiyamotoNagaiPotential(amp=0.140,\
                                             a=0.517,\
                                             b=0.0222)
                                             
thick_disk = potential.MiyamotoNagaiPotential(amp=0.281,\
                                              a=0.3324,\
                                              b=0.0591)
                                              
halo = potential.NFWPotential(amp=4.61,\
                              a=1.477)
                              
bar = potential.SoftenedNeedleBarPotential(amp=0.058,\
                                           a=0.554,\
                                           b=0.111,\
                                           c=0.246,\
                                           omegab=0.736,\
                                           pa=0.41)
                                           
nucleus = potential.PlummerPotential(amp=0.097,\
                                     b=0.036)

# Combined potential
potential = [thin_disk, thick_disk, halo, bar, nucleus]
\end{lstlisting}
\caption{Python code that reproduces the common gravitational potential of the Milky Way in natural units.}
\label{fig:potential_code_2}
\end{figure}



\section{Statistical distance metrics}
\label{sec:metrics}

Simulation of a cluster dissolution produces a sampling of an underlying and unknown probability distribution that describes the likelihood that stars departing the cluster end up at some coordinates in space over time. To test whether two simulations produce different probability distributions, we must be able to measure the statistical distance between the distributions, meaning we must evaluate, with a single number, how different the distributions are from each other. If the distributions were known, this would be a trivial matter; however, we can only simulate a cluster with a finite number of stars which sample the underlying distribution. 

We tested different statistical distance metrics, some frequently used in the astronomical literature, and the Maximum Mean Discrepancy (MMD) \citep{gretton12}, which is commonly used in machine learning \citep{dziugaite15}. We only tested metrics that can compute distances between distributions in at least a two-dimensional space ($XY$ Galactic plane in our case). However, all of the metrics work in higher dimensions as well. We could also have used them on parameter spaces or included stellar properties like masses.  

In the tests of different metrics presented in this section, we compare the values for distances between distributions. In the rest of this work, we treat distances as scores; i.e., lower scores indicate more similar distributions. The distance or score must also be non-negative. Zero indicates the same samples.

\subsection{Maximum mean discrepancy}

MMD is a statistical method to measure differences between sampled probability distributions. MMD is versatile, stable, and able to handle high-dimensional spaces \citep{gretton12}. It can compute distances directly from samples, whereas most alternative methods first infer probability distributions from the samples and then compute distances between them.  MMD computes distances by embedding both probability distributions in a reproducing kernel Hilbert space (also called a feature space) and computing the distances between both distributions from their mean embeddings.  

\subsubsection{Kernel embedding}

Kernel embedding maps a probability distribution into a feature space. It takes a sample $x$ and maps it according to the mapping $\varphi(x)$. An embedding of probability distribution $P$ sampled with many $x$ is the average position of $x$ in the feature space:
\begin{equation}
\label{eq:muP}
    \mu_P = \mathbb{E}_{x \sim P}[\varphi(x)].
\end{equation}
The same can be done for the second probability distribution $Q$:
\begin{equation}
\label{eq:muQ}
    \mu_Q = \mathbb{E}_{y \sim Q}[\varphi(y)].
\end{equation}

The MMD distance between probability distributions $P$ and $Q$ is then
\begin{equation}
    \mathrm{MMD}^2(P, Q) = \left\| \mu_P - \mu_Q \right\|_{\mathcal{H}}^{2},
\end{equation}
where $\mathcal{H}$ marks that the computation is done in the feature space (a Hilbert space). 

\subsubsection{Kernel trick}

$\varphi(x)$ is in practice computed with a kernel trick. A kernel $k(x, x')$ measures the similarity between two points, in the feature space. This is an inner product between mappings $\varphi(x)$ and $\varphi(x')$:
\begin{equation}
    k(x, x') = \langle \varphi(x), \varphi(x') \rangle_{\mathcal{H}}.
\end{equation}
Squared MMD distance can also be written as an inner product
\begin{equation}
    \mathrm{MMD}^2(P, Q) = \left\| \mu_P - \mu_Q \right\|_{\mathcal{H}}^{2}=\langle \mu_P-\mu_Q, \mu_P-\mu_Q \rangle_{\mathcal{H}} = \langle \mu_P, \mu_P \rangle_{\mathcal{H}}
- 2 \langle \mu_P, \mu_Q \rangle_{\mathcal{H}}
+ \langle \mu_Q, \mu_Q \rangle_{\mathcal{H}}.
\end{equation}
Substituting the three terms with the definitions in equations \ref{eq:muP} and \ref{eq:muQ}, we get
\begin{equation}
    \mathrm{MMD}^2(P, Q)
= \mathbb{E}_{x, x' \sim P}[k(x, x')]
- 2\,\mathbb{E}_{x \sim P, y \sim Q}[k(x, y)]
+ \mathbb{E}_{y, y' \sim Q}[k(y, y')]
\end{equation}
We coded the above equation for MMD in \textsc{PyTorch} \citep{paszke19} \footnote{\url{https://github.com/pytorch}, we also adapted code from \url{https://github.com/jindongwang/transferlearning}.}. 

\subsubsection{Kernels}

The core of the MMD algorithm is a kernel mean embedding, so for MMD we also compare different kernels. Here we test the performance of the following kernels.

\paragraph{Gaussian kernel} has a form
\begin{equation}
K(d) = \exp\left(-\frac{d^2}{2\sigma^2}\right),
\end{equation}
where $\sigma$ controls the bandwidth. 

\paragraph{Polynomial kernel} has a form
\begin{equation}
    K(\mathbf{x}, \mathbf{y}) = \left( \frac{\mathbf{x}^\top \mathbf{y}}{\sigma^2} + c \right)^d,
\end{equation}
where $\mathbf{p}$ and $\mathbf{q}$ are vectors consisting of samples that represent each distribution function. Constants $c$ and $d$ define the shape of the kernel, and $\sigma$ acts similarly as bandwidth in the Gaussian kernel. We used $c=1$ and tried various powers of $d$. Figures \ref{fig:metric_test_1k} and \ref{fig:metric_test_4k} shows results with $d=2$.

\paragraph{Matern 3/2 kernel} has a form
\begin{equation}
    K(d) = \left(1 + \frac{\sqrt{3} \, d}{\ell} \right) \exp\left(-\frac{\sqrt{3} \, d}{\ell} \right),
\end{equation}
where $\ell$ controls the scale of the bandwidth of the kernel.

\paragraph{Bump kernel} has a form
\begin{equation}
K(d) = 
\begin{cases}
\exp\left( \frac{1}{\left( d/d_0 \right )^2 -1}\right), & \text{ if } d < d_0, \\
0, & \text{ if } d \geq d_0, 
\end{cases}
\end{equation}
where $d_0$ controls it's bandwidth.

\paragraph{Sinc kernel} has a form
\begin{equation}
    K(d)=\frac{\sin(A \pi d)}{A \pi d},
\end{equation}
where $A$ controls the bandwidth.

\paragraph{Bandwidth} can be a fixed value or can be estimated from the data. In the latter case, it is important that we give higher weights to the samples/stars in the tidal tails and not the core. We want the metrics to be sensitive to changes observed in the tails, even if most stars might reside in the core. We computed a heuristic bandwidth estimate from both sets of samples that are compared in the tests. We produce a combined set of samples and compute a median pairwise Euclidean distance between the elements in the combined set, ignoring stars within the tidal radius. A typical bandwidth used in the tests below is around $\sigma=250\ \mathrm{pc}$.

\subsection{Other metrics}

\paragraph{Earth mover's distance} (EMD) is defined as the minimal cost to transform one distribution into another by moving sample points from the position where they represent the first distribution to the positions where they represent the other distribution \citep{rubner98}. In our test, all points had equal weights. The requirement for the EMD is that the two probability distributions are sampled with the equal number of points. In practice, this can be a decisive limitation, if we wanted to compare clusters of different sizes, although in this work the simulations can be made such that the number of stars is constant. In the end, we did not use this distance, except in the test in this section. The computational complexity of this method is $\mathcal{O}(n^3\log n)$, but can be approximated faster \citep{shirdhonkar08}. We used an exact method here, which is able to compute all distances, for samples with 4000 stars, for all test shown here in several CPU-hours. Using exact EMD for any larger system would be impractical. 

\paragraph{Kullback-Leibler divergence} (KLD) measures how well one distribution approximates a true distribution \citep{kullback51}.  The main drawback of this method is that it is not symmetric. If we compare two distributions, it matters which one we pick to represent the `true' distribution. In practice we do not know this. In the test in this section, we always use the same distribution against which we compare modified and distorted distributions, so the choice of `true' distribution is obvious. Elsewhere in the paper, we compute a symmetric KLD:
\begin{equation}
    D(P,Q)=D_\mathrm{KL}(P||Q)+D_\mathrm{KL}(Q||P).
\end{equation}
KLD also does not satisfy the triangle inequality, which should not be an issue for the way we use the KLD in this work.

\paragraph{Kolmogorov-Smirnov statistics} measures the difference between two 1D distributions. It is a widely used metric, so here we test it's performance in a sliced version, where a similarity of two 2D distributions is defined as the sum of Kolmogorov-Smirnov statistics when applied to each dimension independently. We also tried a 2D Smirnov-Kolmogorov test \citep{peacock83, fasano87}, but it did not perform as well as a simple sliced Kolmogorov-Smirnov test. We speculate that this is because our data has a strong principal axis, which makes a 1D test as sensitive as it can be. 

\subsection{Toy model}

\begin{figure*}
    \centering
    \includegraphics[width=0.99\textwidth]{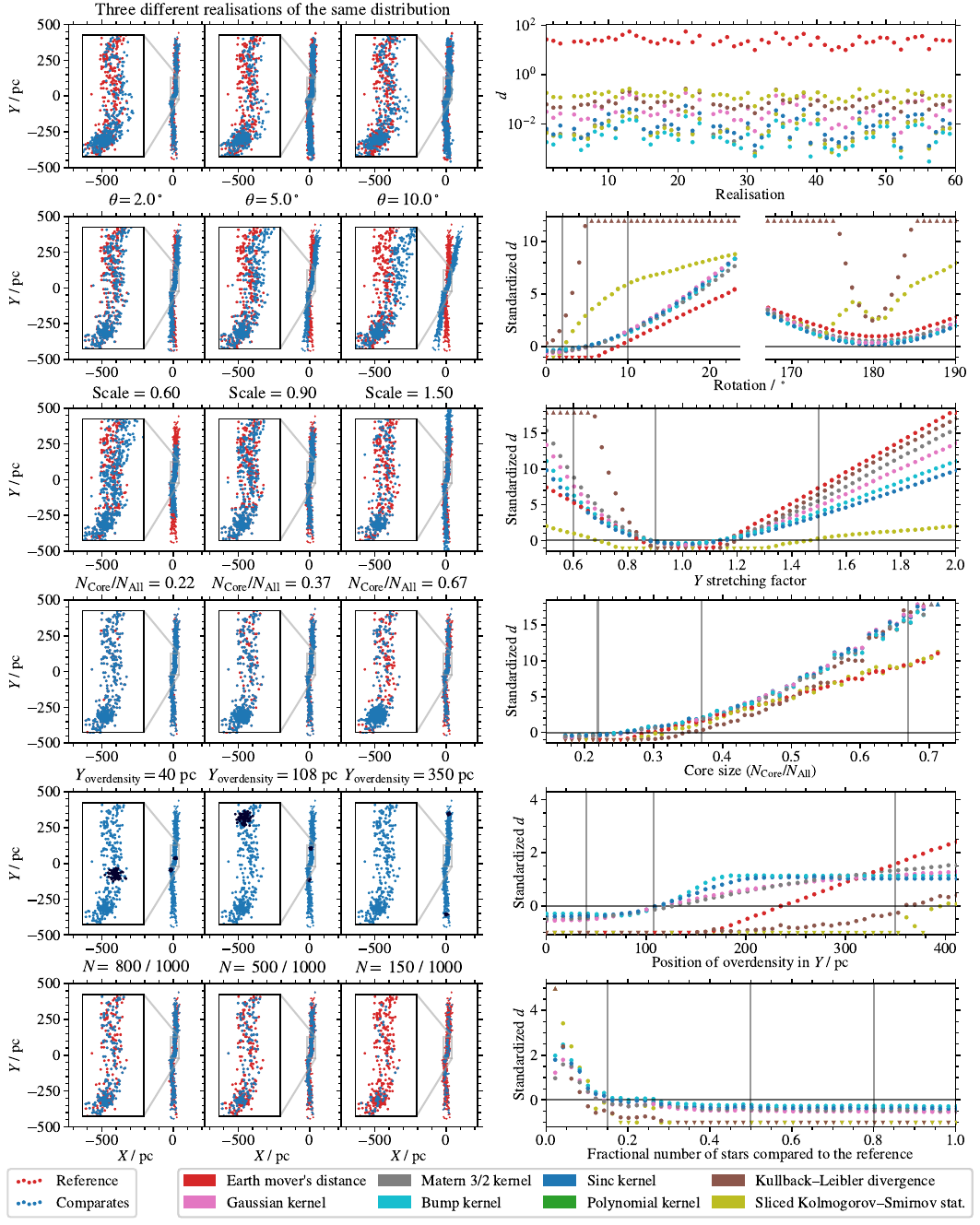}
    \caption{Performance of various metrics to measure the differences between two distributions of 1000 stars in a dissolving cluster. Left column of panels illustrates the variations of structures and the right column the sensitivity of metrics when trying to detect the variations. Top to bottom, we show the sensitivity to: statistical noise, rotation of distributions, scaling of tidal tails, density of the core, position of additional overdensities (marked in black) inside the tails, and the difference in the number of samples. Top-right panels shows the distances, and other panels standardized distances based on the mean and scatter from the top panel.}
    \label{fig:metric_test_1k}
\end{figure*}

\begin{figure*}
    \centering
    \includegraphics[width=0.99\textwidth]{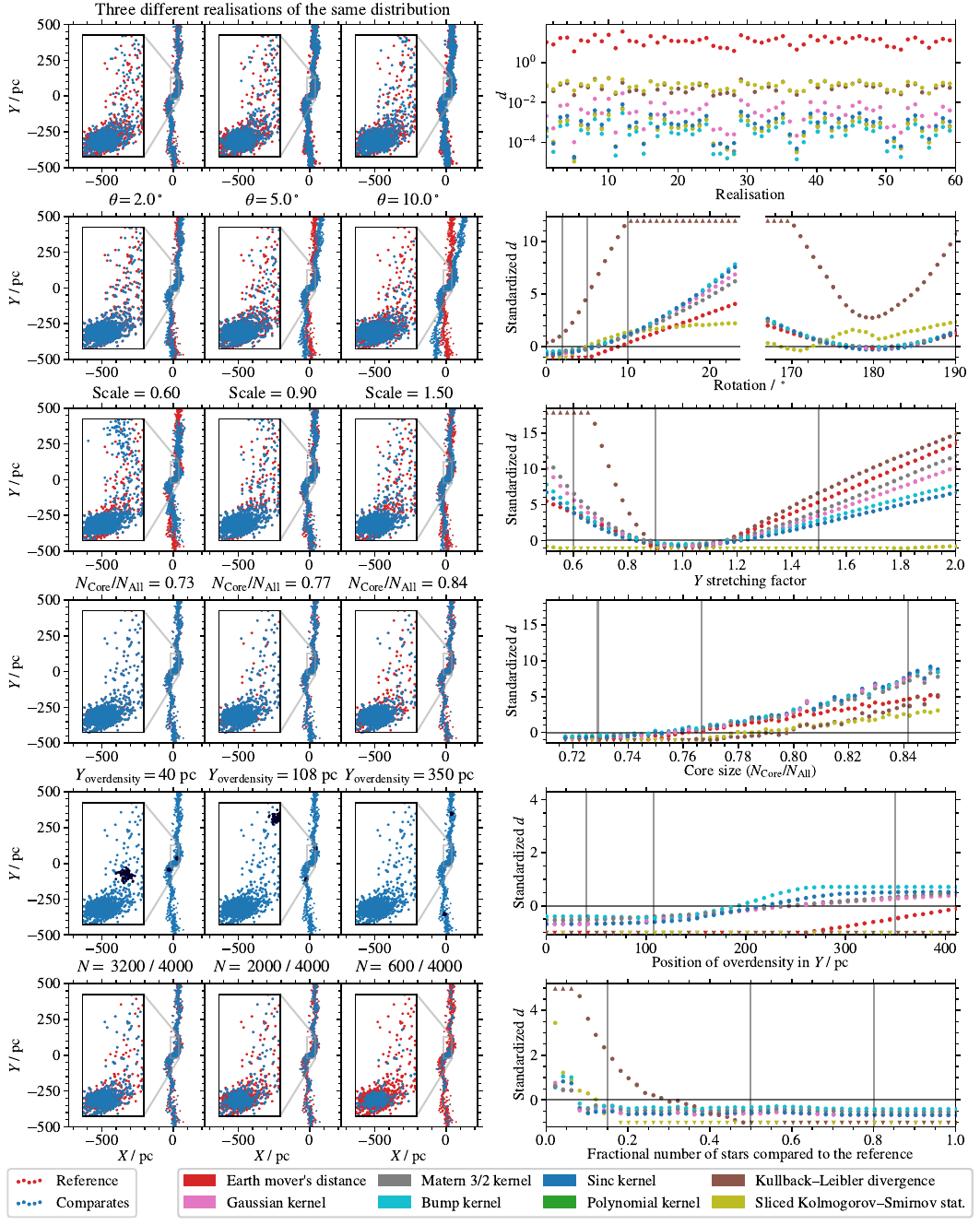}
    \caption{Same as Figure \ref{fig:metric_test_1k} but for a cluster of 4000 stars.}
    \label{fig:metric_test_4k}
\end{figure*}

 All tests presented below use a toy model -- one of our simulations that we artificially modified and distorted to simulate a number of features to which our metrics and kernels should be sensitive. We used a simulation of Hyades (age of $577\ \mathrm{Myr}$) as they dissolve in the common gravitational potential of the Galaxy. The simulated positions of stars in the $XY$ plane were then artificially modified to test different cases of feature detection by the above metrics. Six tests were performed and are described below. The results are visualised in Figures \ref{fig:metric_test_1k} and \ref{fig:metric_test_4k} and summarised in Section \ref{sec:metrics_conclusions}. 

\paragraph{Consistency}

We repeated the simulation multiple times, always with the same initial conditions, same gravitational potential, but with randomised population of the initial King's distribution of stars. The result was several realisations that sample the same distribution. Distances between different realisations serve as a baseline for other tests. For each metric we computed a median and a standard deviation of distances between the first realisation and all subsequent realisations. If the distance computed in other tests is smaller than the median, the variation of a parameter in such test is not detected. If the distance is larger than the median, we can estimate to how many standard deviations some metric is sensitive when some parameter described below is varied. Note that the scatter of distances between different realisations is not normally distributed but is concentrated closer to the median. In our tests, sensitivity, when expressed in standard deviations, is hence underestimated.

\paragraph{Rotation}

We rotated the simulated distribution of stars around the cluster centre and tested at which angle the metrics detect the change. 

\paragraph{Scale}

We rescaled the cluster in $Y$ direction using the formula
\begin{equation}
    Y_\mathrm{rescaled}=Y\left[\frac{s-1}{1+\exp(-0.6 (|y|-15.0))}+1\right],
\end{equation}
where $s$ is a scale factor, which we varied between $0.5<s<2.0$. Using this formula, the tails beyond $15\ \mathrm{pc}$ from the core get stretched, while the core keeps the same shape, and there is a smooth transition between both regimes. 

\paragraph{Concentration}

We varied the ratio of stars in the core to those in the tails. We took some fraction of stars from the tails and moved them into the core, effectively making the tails less populated by stars. The total number of stars always remained the same. 

\paragraph{Density profile of the tails}

We added a small clump of stars (50 stars) to the tails, symmetrically at different distances from the cluster core. Because the star density in the tails is dropping with distance from the core, this additional overdensity becomes more pronounced at larger distances.  

\paragraph{Cluster sizes}

A good metric must be able to measure the distance between the underling probability distributions, even if they are sampled with a different number of points. We test whether the distance increases from zero when we randomly remove stars from the initial distribution and computing the distance between the original and sparser sampling. We find that all metrics perform well when no more than 60\% of the points are removed. 

\subsection{Conclusions on best suited similarity metrics}
\label{sec:metrics_conclusions}

The Kolmogorov-Smirnov statistic and the Earth mover's distance showed obvious drawbacks against other metrics tested here. Kolmogorov-Smirnov statistic was not sensitive to scaling and additional overdensities, and was least sensitive in all other tests. Earth's mover distance performed slightly better, but was competitive with other metrics only in the scaling test. KLD showed by far the highest sensitivity to rotation and the highest sensitivity to scaling. In other tests the MMD performed best. We focused on the kernels that best work with our data. The outlier was the polynomial kernel, which is much less sensitive to scaling. It was only able to detect a contraction of 25\% or expansion of 50\%. In Figures \ref{fig:metric_test_1k} and  \ref{fig:metric_test_4k} we only show a polynomial kernel of the second order, but increasing the order did not improve the performance. Bump and sinc kernels performed best, over-performing Gaussian kernel and Matern 3/2 kernel particularly in the test with an added overdensity. They were able to detect a small additional overdensity containing only 2\% of all the stars in the cluster at a distance smaller than a typical distance of an epicyclic overdensity from the core in real clusters (100 -- 150 pc). 

Because no single metric was able to perform well in all tests, we chose to use both the KLD and the MMD metric with a bump kernel to quantify the differences between cluster simulations produced in this work. KLD is able to detect rotations larger than $\sim 1^\circ$ and the MMD metric with a bump kernel is able to distinguish a change in core density of less than 10\%, and small changes in the density profile of the tails at the level of the density variations inside and outside the epicyclic overdensities. Both KLD and MMD are sensitive to scaling. The performance of MMD and KLD can be further improved by disregarding the stars in the core of the cluster.  

\section{Grid of orbits}

\begin{figure*}
    \includegraphics[width=\textwidth]{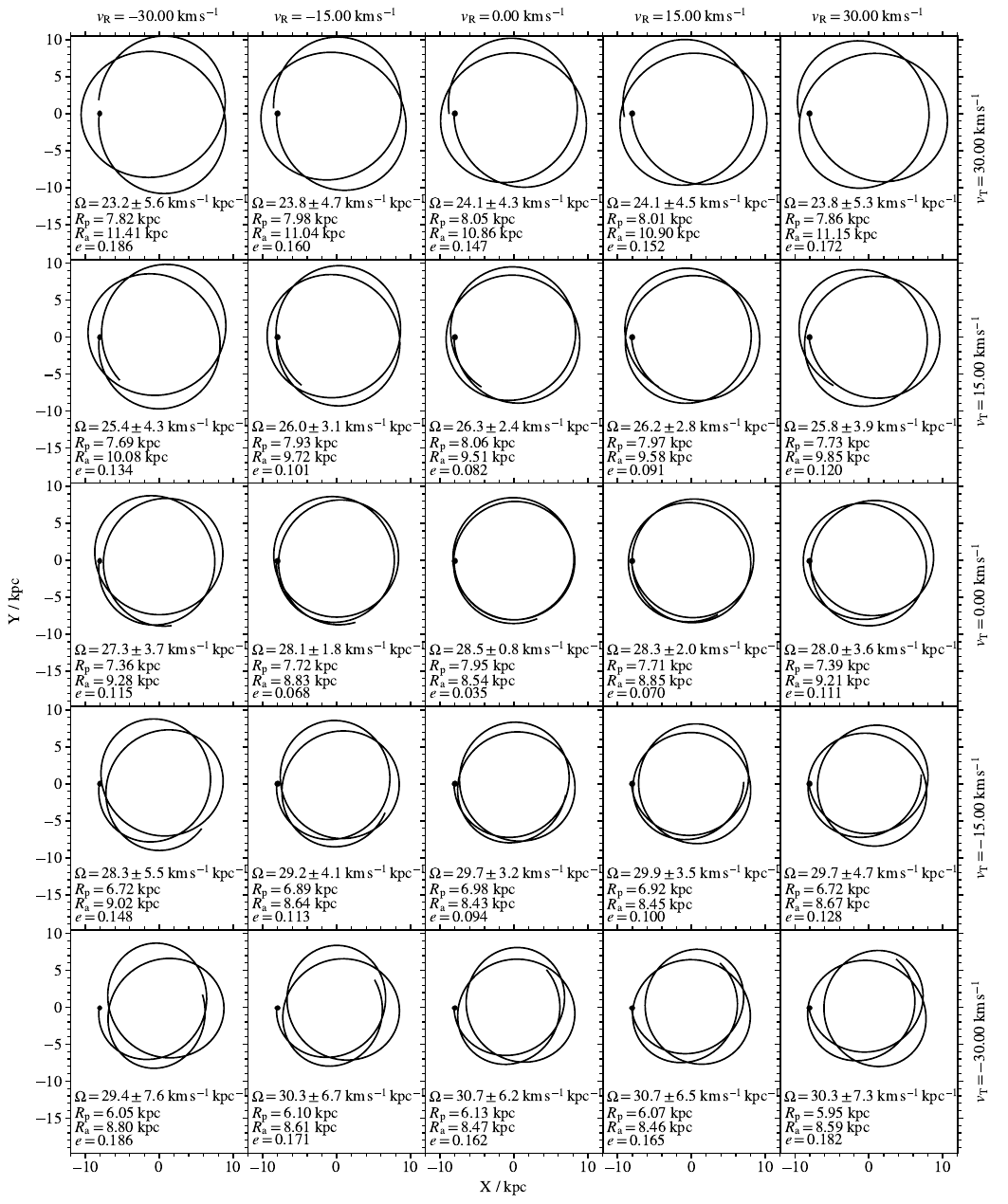}
    \caption{Orbits of cluster centres integrated for the past $500\ \mathrm{Myr}$ in the common potential. Grid of panels show orbits for 25 different combinations of current $v_\mathrm{R}$ and $v_\mathrm{T}$ velocities. The final positions are marked with a dot, and the positions of clusters $500\ \mathrm{Myr}$ ago are at the opened ends of the solid lines. Values in each panel show the mean angular velocity over the plotted orbit ($\Omega$) with one standard deviation of the angular velocity variability given after the $\pm$ sign. Below we list the pericentre ($R_\mathrm{p}$) and apocentre ($R_\mathrm{a}$) of the orbit, and its eccentricity ($e$).}
    \label{fig:typical_orbits}
\end{figure*}

We plotted sensitivities of tidal tails to perturbations and variations of the common galactic potential in Section \ref{sec:sensitivity} for a grid of orbital parameters of synthetic clusters. Figure \ref{fig:typical_orbits} shows the shape of the orbits for the grid used in Section \ref{sec:sensitivity}. Each panel shows an orbit integrated back in time for $500\ \mathrm{Myr}$ from the initial conditions matching the position of the Sun and with the velocity $v_\mathrm{z}=0$, and $v_\mathrm{R}$ and $v_\mathrm{T}$ marked on the edge of the grid. We also marked angular velocity, pericentre, apocentre and eccentricity of each orbit in the grid. Orbits in perturbed potentials can differ from those shown in Figure \ref{fig:typical_orbits} but retain the same general characteristics. 

\section{List of real open clusters used in simulations}
\label{sec:clusters_table_appendix}
In Section \ref{sec:real_clusters} we described simulations of the dissolution of real open clusters. Table \ref{tab:clusters} provides information on the clusters used in these simulations. Along with the names of the clusters, we list their ages, current positions and velocities, pericentres, apocentres, and eccentricities of orbits, maximum distances from the Galactic plane and mean angular velocity with one standard deviation of the angular velocity variability given after the $\pm$ sign. Cluster ages are taken from \citep{hunt24}, current positions and velocities in the cartesian coordinate system are computed from celestial positions and velocities given in \citep{hunt24}, and orbital parameters are obtained by integrating cluster orbits back in time for 1 Gyr. Orbital parameters are calculated for the common potential and can vary for different pattern speeds of the bar.
\setlength{\tabcolsep}{3.13pt}
\setlength\LTcapwidth{\linewidth}
\renewcommand{\arraystretch}{0.98}
\begin{longtable}{l c c c c c c c c c c c c}
\caption{Real open clusters used in simulations. Along with a cluster's name, the table lists its age, current position and velocity, pericentre, apocentre, orbital eccentricity, maximum distance from the Galactic plane and mean angular velocity along the orbit.} \\
\label{tab:clusters}

Name & Age & $X$ & $Y$ & $Z$ & $U$ & $V$ & $W$ & $R_p$ & $R_a$ & $e$ & $z_{\rm max}$ & $\Omega$ \\ \hline
& Myr & kpc & kpc & kpc & $\mathrm{km\,s^{-1}}$ & $\mathrm{km\,s^{-1}}$ & $\mathrm{km\,s^{-1}}$ & kpc & kpc & & kpc & $\mathrm{km\,s^{-1}\,kpc^{-1}}$ \\ \hline\hline
\endfirsthead

\multicolumn{8}{c}%
{{\tablename\ \thetable{} -- continued.}} \\
Name & Age & $X$ & $Y$ & $Z$ & $U$ & $V$ & $W$ & $R_p$ & $R_a$ & $e$ & $z_{\rm max}$ & $\Omega$ \\ \hline
& Myr & kpc & kpc & kpc & $\mathrm{km\,s^{-1}}$ & $\mathrm{km\,s^{-1}}$ & $\mathrm{km\,s^{-1}}$ & kpc & kpc & & kpc & $\mathrm{km\,s^{-1}\,kpc^{-1}}$ \\ \hline\hline
\endhead

\endfoot

\hline 
\endlastfoot

ASCC 23 & 291.2 & -8.6925 & 0.1267 & 0.1723 & 25.29 & 239.81 & 6.91 & 8.526 & 9.685 & 0.099 & 0.237 & 25.0$\pm$2.2\\
ASCC 87 & 431.0 & -7.3804 & -0.0633 & 0.1240 & 12.18 & 232.03 & 3.72 & 6.947 & 7.980 & 0.072 & 0.141 & 31.3$\pm$2.5\\
ASCC 90 & 409.2 & -7.5542 & -0.0570 & 0.0003 & 18.30 & 236.50 & 7.22 & 7.105 & 8.482 & 0.092 & 0.119 & 29.9$\pm$3.0\\
ASCC 123 & 1064.5 & -8.2366 & 0.4372 & -0.0107 & 6.91 & 233.70 & -9.30 & 8.098 & 8.643 & 0.033 & 0.166 & 27.9$\pm$0.8\\
Alessi 2 & 263.8 & -8.6536 & 0.2789 & 0.0888 & 21.22 & 240.54 & 2.87 & 8.659 & 9.466 & 0.077 & 0.110 & 25.2$\pm$1.5\\
Alessi 3 & 625.5 & -8.1779 & -0.2608 & -0.0522 & -5.89 & 250.44 & 2.19 & 8.121 & 9.703 & 0.089 & 0.073 & 25.9$\pm$2.8\\
Alessi 6 & 365.5 & -7.5307 & -0.6207 & -0.0640 & -32.64 & 222.27 & 5.98 & 6.786 & 7.853 & 0.073 & 0.120 & 32.2$\pm$2.3\\
Alessi 9 & 411.2 & -7.9237 & -0.0554 & -0.0113 & 1.91 & 245.52 & -3.72 & 7.871 & 8.932 & 0.070 & 0.065 & 27.5$\pm$2.0\\
Alessi 22 & 671.8 & -8.2269 & 0.3034 & -0.1282 & 29.12 & 237.67 & -7.36 & 7.825 & 9.240 & 0.085 & 0.195 & 27.2$\pm$2.6\\
Alessi 62 & 425.4 & -7.7594 & 0.4774 & 0.1120 & 23.31 & 255.34 & 8.11 & 7.775 & 9.719 & 0.112 & 0.196 & 26.5$\pm$3.6\\
Alessi 116 & 1094.8 & -7.7375 & 0.7894 & -0.1859 & 65.70 & 245.66 & 0.33 & 7.261 & 10.058 & 0.161 & 0.223 & 26.6$\pm$5.4\\
Alessi 191 & 958.0 & -7.5327 & 0.7666 & 0.1979 & -2.10 & 244.18 & 17.15 & 7.187 & 8.656 & 0.093 & 0.373 & 29.5$\pm$3.1\\
COIN-Gaia 11 & 339.1 & -8.7321 & 0.1921 & -0.0423 & 11.98 & 224.97 & 3.99 & 8.029 & 9.107 & 0.063 & 0.088 & 27.2$\pm$2.0\\
COIN-Gaia 25 & 574.5 & -8.8997 & -0.1333 & 0.0208 & 27.58 & 240.04 & 1.40 & 8.312 & 10.478 & 0.134 & 0.037 & 24.7$\pm$3.4\\
CWNU 41 & 628.5 & -7.6322 & -0.5217 & -0.1172 & -10.62 & 231.16 & -4.39 & 7.285 & 8.030 & 0.049 & 0.137 & 30.3$\pm$1.4\\
CWNU 61 & 318.5 & -7.9358 & -0.4915 & -0.0912 & -6.30 & 226.21 & 4.83 & 7.344 & 8.325 & 0.068 & 0.129 & 29.6$\pm$2.1\\
CWNU 418 & 358.5 & -8.9347 & -0.3367 & -0.0064 & 4.83 & 229.05 & -4.88 & 8.289 & 9.487 & 0.073 & 0.096 & 26.1$\pm$2.0\\
CWNU 509 & 874.7 & -8.4248 & -0.8612 & 0.2258 & -17.32 & 241.32 & -3.43 & 8.324 & 9.559 & 0.069 & 0.257 & 26.1$\pm$1.8\\
CWNU 1095 & 254.3 & -8.2362 & 0.1042 & 0.2107 & 26.92 & 248.45 & 4.63 & 8.169 & 9.750 & 0.117 & 0.255 & 25.2$\pm$3.1\\
Casado-Alessi 1 & 748.5 & -8.4927 & 0.5645 & -0.1381 & 0.33 & 225.90 & -11.72 & 7.736 & 9.020 & 0.077 & 0.265 & 28.0$\pm$2.4\\
ESO 123-26 & 361.2 & -8.0739 & -0.8869 & -0.2388 & -35.90 & 226.67 & 10.74 & 7.704 & 8.293 & 0.048 & 0.337 & 29.0$\pm$1.3\\
Ferrero 11 & 351.0 & -8.8063 & -0.3656 & -0.0848 & 4.09 & 230.32 & -11.96 & 8.235 & 9.320 & 0.071 & 0.256 & 26.3$\pm$1.9\\
Gulliver 20 & 255.1 & -7.8052 & 0.2582 & 0.1153 & 10.19 & 229.41 & -2.29 & 7.442 & 8.123 & 0.049 & 0.127 & 30.0$\pm$1.4\\
Gulliver 21 & 281.9 & -8.4604 & -0.5315 & -0.0674 & -20.05 & 219.35 & 2.56 & 7.497 & 8.705 & 0.089 & 0.087 & 29.0$\pm$2.5\\
HSC 95 & 313.4 & -7.9324 & 0.0216 & 0.0529 & -4.65 & 246.44 & -10.33 & 7.916 & 8.868 & 0.069 & 0.201 & 27.5$\pm$1.9\\
HSC 147 & 1044.9 & -7.6057 & 0.1014 & -0.1233 & 22.92 & 226.09 & 2.17 & 6.785 & 8.265 & 0.098 & 0.136 & 30.7$\pm$3.5\\
HSC 170 & 1769.2 & -7.8750 & 0.0593 & -0.1153 & 0.33 & 223.86 & 6.26 & 7.148 & 8.125 & 0.064 & 0.164 & 30.4$\pm$2.1\\
HSC 252 & 2106.2 & -7.5539 & 0.2743 & -0.1040 & -36.81 & 228.51 & -7.99 & 6.341 & 8.600 & 0.151 & 0.167 & 31.2$\pm$5.9\\
HSC 442 & 275.9 & -8.0250 & 0.1316 & 0.0231 & 3.05 & 225.86 & 2.50 & 7.428 & 8.282 & 0.058 & 0.046 & 29.8$\pm$1.8\\
HSC 507 & 529.8 & -8.0617 & 0.1166 & 0.0207 & 5.39 & 231.38 & -0.18 & 7.804 & 8.461 & 0.047 & 0.022 & 28.7$\pm$1.2\\
HSC 590 & 318.9 & -8.0965 & 0.0962 & 0.0845 & 14.38 & 240.83 & 6.32 & 7.914 & 8.913 & 0.069 & 0.144 & 27.6$\pm$1.9\\
HSC 900 & 380.0 & -8.4393 & 0.7432 & -0.0179 & 12.96 & 230.69 & 1.16 & 8.133 & 8.815 & 0.041 & 0.028 & 27.7$\pm$1.1\\
HSC 1029 & 711.0 & -8.6587 & 0.6785 & -0.1861 & 52.12 & 233.26 & 1.27 & 7.891 & 10.162 & 0.126 & 0.209 & 25.7$\pm$4.0\\
HSC 1074 & 1261.6 & -8.6938 & 0.6399 & 0.1026 & 16.27 & 199.65 & 18.47 & 6.482 & 9.048 & 0.165 & 0.411 & 29.9$\pm$6.5\\
HSC 1142 & 430.5 & -8.3115 & 0.1457 & 0.0749 & 9.36 & 239.42 & -9.73 & 8.226 & 9.073 & 0.050 & 0.204 & 26.7$\pm$1.3\\
HSC 1179 & 334.6 & -8.6075 & 0.3150 & -0.0166 & 29.29 & 244.15 & 2.18 & 8.517 & 9.858 & 0.105 & 0.047 & 25.2$\pm$2.5\\
HSC 1195 & 253.4 & -8.2009 & 0.0452 & -0.0312 & 20.19 & 241.80 & 2.67 & 8.125 & 9.218 & 0.089 & 0.060 & 26.2$\pm$2.2\\
HSC 1513 & 263.2 & -8.8343 & -0.1441 & 0.1661 & 31.72 & 233.81 & 3.48 & 8.123 & 9.946 & 0.135 & 0.197 & 24.5$\pm$3.4\\
HSC 1575 & 360.8 & -8.6688 & -0.1804 & 0.0232 & 8.82 & 234.49 & 5.78 & 8.271 & 9.299 & 0.068 & 0.112 & 26.3$\pm$1.8\\
HSC 1658 & 524.8 & -8.4342 & -0.1806 & -0.0427 & -5.14 & 238.38 & -1.55 & 8.418 & 9.076 & 0.038 & 0.051 & 26.7$\pm$1.0\\
HSC 1724 & 1473.3 & -8.2453 & -0.0966 & 0.0364 & 3.91 & 228.04 & 5.21 & 7.787 & 8.668 & 0.054 & 0.097 & 28.5$\pm$1.6\\
HSC 1791 & 2638.8 & -8.5620 & -0.4750 & 0.0004 & -7.77 & 227.42 & 6.63 & 8.055 & 8.935 & 0.052 & 0.122 & 27.4$\pm$1.4\\
HSC 1849 & 432.5 & -8.2834 & -0.2213 & 0.0048 & -18.09 & 229.15 & -3.36 & 7.806 & 8.651 & 0.060 & 0.059 & 28.5$\pm$1.6\\
HSC 1986 & 614.8 & -8.1696 & -0.1108 & -0.0799 & 10.05 & 250.30 & -9.33 & 8.021 & 9.827 & 0.101 & 0.201 & 25.8$\pm$3.2\\
HSC 2082 & 372.7 & -8.2727 & -0.7756 & -0.0255 & -33.41 & 225.49 & 12.91 & 7.756 & 8.532 & 0.062 & 0.248 & 28.7$\pm$1.6\\
HSC 2106 & 312.8 & -8.2362 & -0.7953 & 0.0270 & -27.45 & 221.94 & -2.72 & 7.530 & 8.446 & 0.070 & 0.056 & 29.0$\pm$1.9\\
HSC 2232 & 255.4 & -8.0283 & -0.9904 & 0.0541 & -52.80 & 234.00 & 6.53 & 7.699 & 8.872 & 0.088 & 0.127 & 28.3$\pm$2.5\\
HSC 2251 & 534.5 & -8.0729 & -0.3978 & 0.0882 & 11.17 & 252.91 & 5.83 & 7.807 & 10.104 & 0.128 & 0.151 & 26.1$\pm$4.1\\
HSC 2278 & 570.6 & -8.1043 & -0.1001 & 0.0908 & 8.11 & 242.45 & 7.45 & 7.941 & 9.089 & 0.070 & 0.166 & 27.1$\pm$2.0\\
HSC 2283 & 373.2 & -8.1077 & -0.0792 & -0.0470 & 30.14 & 246.43 & -0.87 & 7.658 & 9.807 & 0.135 & 0.056 & 26.4$\pm$4.1\\
HSC 2327 & 765.5 & -8.0696 & -0.2153 & -0.0034 & -21.06 & 227.04 & 8.18 & 7.432 & 8.343 & 0.058 & 0.141 & 29.5$\pm$1.9\\
HSC 2483 & 395.2 & -7.9577 & -0.3139 & -0.1161 & -7.64 & 227.18 & 3.48 & 7.407 & 8.225 & 0.052 & 0.136 & 29.8$\pm$1.7\\
HSC 2645 & 608.4 & -7.6836 & -0.4594 & -0.1298 & -18.32 & 244.71 & 0.41 & 7.689 & 8.651 & 0.059 & 0.141 & 28.5$\pm$1.8\\
HSC 2745 & 265.9 & -7.5632 & -0.3121 & 0.0278 & -11.13 & 229.13 & -6.44 & 7.188 & 7.749 & 0.042 & 0.107 & 31.2$\pm$1.1\\
HSC 2819 & 381.0 & -7.1619 & -0.3612 & 0.0155 & -25.92 & 232.32 & -2.42 & 6.834 & 7.635 & 0.062 & 0.041 & 32.4$\pm$1.8\\
HSC 2838 & 599.0 & -7.6360 & -0.1576 & -0.0850 & -18.16 & 193.29 & -2.06 & 5.089 & 7.970 & 0.230 & 0.092 & 36.4$\pm$9.7\\
HSC 2873 & 304.0 & -7.6771 & -0.0951 & 0.0374 & -1.20 & 227.58 & -2.74 & 7.172 & 7.943 & 0.054 & 0.060 & 30.7$\pm$1.7\\
HXHWL 5 & 465.9 & -8.6633 & -0.6193 & 0.0383 & 2.69 & 239.93 & 4.69 & 8.365 & 9.914 & 0.093 & 0.100 & 25.2$\pm$2.5\\
IC 4651 & 1665.1 & -7.2741 & -0.3073 & -0.1065 & -23.36 & 232.68 & 8.24 & 6.964 & 7.723 & 0.052 & 0.179 & 31.8$\pm$1.6\\
IC 4756 & 836.9 & -7.7474 & 0.2752 & 0.0633 & -1.47 & 224.57 & -1.70 & 6.994 & 8.042 & 0.070 & 0.073 & 31.1$\pm$2.3\\
LP 2309 & 352.1 & -7.5113 & -0.7165 & -0.0094 & -2.84 & 218.98 & 7.29 & 6.389 & 8.084 & 0.117 & 0.119 & 31.9$\pm$4.2\\
Loden 995 & 391.7 & -7.4869 & -0.7657 & -0.0316 & -17.73 & 228.91 & -2.82 & 7.089 & 7.879 & 0.053 & 0.053 & 31.1$\pm$1.7\\
Mamajek 4 & 312.0 & -7.7133 & -0.1165 & -0.1085 & -32.05 & 217.38 & -7.51 & 6.396 & 8.207 & 0.124 & 0.161 & 32.3$\pm$4.7\\
Melotte 25 & 576.8 & -8.1657 & 0.0001 & 0.0031 & -28.80 & 225.88 & 6.03 & 7.172 & 8.700 & 0.096 & 0.105 & 29.4$\pm$3.4\\
Melotte 111 & 654.2 & -8.1290 & -0.0056 & 0.1056 & 10.46 & 239.93 & 7.40 & 7.947 & 8.886 & 0.064 & 0.174 & 27.6$\pm$1.7\\
NGC 752 & 1420.6 & -8.4133 & 0.2721 & -0.1500 & -3.52 & 225.48 & -11.35 & 7.705 & 8.874 & 0.071 & 0.261 & 28.2$\pm$2.1\\
NGC 1662 & 345.5 & -8.4963 & -0.0507 & -0.1241 & 26.82 & 245.47 & 8.42 & 8.212 & 9.974 & 0.127 & 0.232 & 25.5$\pm$3.4\\
NGC 1758 & 305.8 & -8.9657 & 0.0124 & -0.1327 & 10.27 & 226.59 & 8.95 & 8.260 & 9.351 & 0.069 & 0.238 & 26.4$\pm$2.0\\
NGC 1901 & 282.7 & -8.0681 & -0.3421 & -0.2103 & -12.13 & 241.81 & 5.64 & 8.076 & 8.766 & 0.053 & 0.259 & 27.3$\pm$1.2\\
NGC 2281 & 274.2 & -8.6066 & 0.0434 & 0.1700 & -11.16 & 231.66 & 0.57 & 8.219 & 9.022 & 0.048 & 0.175 & 27.2$\pm$1.5\\
NGC 2358 & 419.1 & -8.6855 & -0.6981 & -0.0134 & -8.10 & 226.91 & 2.78 & 8.067 & 9.157 & 0.065 & 0.054 & 26.9$\pm$1.9\\
NGC 2423 & 1182.4 & -8.6988 & -0.6998 & 0.0790 & 9.70 & 222.39 & -1.08 & 7.466 & 9.506 & 0.120 & 0.085 & 27.3$\pm$4.1\\
NGC 2527 & 819.9 & -8.3746 & -0.5698 & 0.0420 & -28.47 & 218.64 & 7.05 & 7.329 & 8.639 & 0.083 & 0.131 & 29.0$\pm$2.9\\
NGC 2548 & 379.3 & -8.6043 & -0.5344 & 0.2205 & 3.27 & 242.59 & 8.47 & 8.439 & 9.686 & 0.096 & 0.308 & 25.5$\pm$2.3\\
NGC 2632 & 346.2 & -8.2610 & -0.0677 & 0.1197 & -29.05 & 225.38 & -1.51 & 7.283 & 8.734 & 0.092 & 0.130 & 29.1$\pm$3.3\\
NGC 2682 & 1688.3 & -8.6980 & -0.4146 & 0.4650 & -33.99 & 221.07 & -10.98 & 7.683 & 9.301 & 0.095 & 0.578 & 27.5$\pm$3.0\\
NGC 5822 & 1039.9 & -7.4898 & -0.5013 & 0.0704 & -31.45 & 235.15 & 2.02 & 7.211 & 8.118 & 0.059 & 0.078 & 30.4$\pm$1.9\\
NGC 6134 & 872.4 & -7.1311 & -0.4639 & 0.0144 & -15.32 & 248.24 & -15.93 & 7.046 & 8.396 & 0.088 & 0.270 & 30.1$\pm$3.0\\
NGC 6281 & 274.0 & -7.6133 & -0.1102 & 0.0376 & 5.16 & 235.98 & 5.40 & 7.475 & 8.225 & 0.059 & 0.094 & 29.8$\pm$1.6\\
NGC 6425 & 443.5 & -7.1437 & -0.0350 & -0.0088 & 6.39 & 247.99 & -11.19 & 7.043 & 8.358 & 0.085 & 0.181 & 30.5$\pm$2.9\\
NGC 6568 & 555.6 & -7.0993 & 0.1694 & -0.0120 & -30.89 & 233.50 & 3.39 & 6.309 & 8.095 & 0.129 & 0.055 & 32.8$\pm$4.8\\
NGC 6618 & 390.1 & -7.5377 & 0.1588 & 0.0135 & -0.65 & 230.26 & -3.69 & 7.168 & 7.807 & 0.043 & 0.060 & 31.2$\pm$1.2\\
NGC 6633 & 472.4 & -7.8103 & 0.2268 & 0.0766 & -7.86 & 228.10 & 0.65 & 7.178 & 8.150 & 0.063 & 0.079 & 30.6$\pm$2.0\\
NGC 6800 & 284.3 & -7.6002 & 0.8767 & 0.0903 & 6.06 & 229.36 & 3.10 & 6.935 & 8.159 & 0.081 & 0.102 & 30.9$\pm$2.9\\
NGC 6940 & 769.7 & -7.7766 & 0.9413 & -0.1059 & 54.25 & 237.05 & -12.64 & 7.434 & 8.986 & 0.103 & 0.262 & 28.2$\pm$3.2\\
NGC 6991 & 989.5 & -8.0962 & 0.5557 & 0.0358 & -14.18 & 234.77 & 10.66 & 7.380 & 9.184 & 0.109 & 0.200 & 28.5$\pm$3.6\\
NGC 6996 & 318.5 & -8.0537 & 0.8597 & 0.0133 & 44.22 & 224.94 & 2.85 & 7.456 & 8.771 & 0.081 & 0.051 & 28.8$\pm$2.9\\
OCSN 3 & 705.2 & -7.9742 & 0.0297 & 0.0097 & 3.42 & 227.29 & -5.47 & 7.382 & 8.276 & 0.057 & 0.092 & 29.7$\pm$1.9\\
OCSN 49 & 469.9 & -8.3274 & 0.0475 & 0.0611 & 10.70 & 240.98 & 0.85 & 8.194 & 9.296 & 0.063 & 0.067 & 26.5$\pm$1.8\\
PHOC 40 & 375.4 & -7.9021 & 0.2407 & 0.0324 & 37.18 & 250.85 & 4.13 & 7.615 & 9.763 & 0.137 & 0.082 & 26.6$\pm$4.2\\
Ruprecht 135 & 283.7 & -7.0869 & 0.3046 & 0.1369 & 3.03 & 239.99 & -2.91 & 7.032 & 7.699 & 0.045 & 0.155 & 31.9$\pm$1.4\\
Ruprecht 145 & 568.0 & -7.5611 & 0.1657 & -0.0634 & -43.96 & 228.69 & -10.36 & 6.300 & 8.724 & 0.161 & 0.185 & 31.2$\pm$6.3\\
Ruprecht 147 & 888.7 & -7.8466 & 0.1058 & -0.0469 & 61.97 & 226.61 & -16.62 & 6.472 & 9.560 & 0.193 & 0.338 & 29.1$\pm$7.5\\
Ruprecht 163 & 374.3 & -7.7625 & -0.8407 & -0.0952 & -2.92 & 221.76 & -4.19 & 6.752 & 8.369 & 0.111 & 0.119 & 30.4$\pm$4.0\\
Stock 1 & 258.1 & -7.9223 & 0.3491 & 0.0359 & -1.69 & 232.35 & -2.67 & 7.549 & 8.209 & 0.043 & 0.059 & 29.6$\pm$1.4\\
Stock 2 & 290.2 & -8.3763 & 0.2690 & 0.0112 & -17.19 & 227.98 & -6.26 & 7.515 & 8.892 & 0.098 & 0.117 & 28.4$\pm$3.0\\
TPK 1 & 273.4 & -8.2953 & 0.4667 & -0.0997 & 10.43 & 231.15 & 5.61 & 8.089 & 8.602 & 0.033 & 0.149 & 28.0$\pm$1.0\\
Teutsch J0450.9+5209 & 283.4 & -8.3907 & 0.1255 & 0.0471 & -11.28 & 227.44 & 3.92 & 7.721 & 8.737 & 0.072 & 0.083 & 28.4$\pm$2.0\\
Theia 47 & 406.0 & -8.7910 & 0.3579 & 0.0304 & 5.65 & 238.80 & -0.28 & 8.773 & 9.328 & 0.038 & 0.033 & 25.6$\pm$0.7\\
Theia 89 & 685.8 & -7.5648 & -0.4658 & -0.0188 & 9.78 & 227.93 & -4.84 & 6.764 & 8.342 & 0.107 & 0.080 & 30.8$\pm$3.9\\
Theia 180 & 314.9 & -7.7643 & 0.4530 & -0.1812 & 1.63 & 243.20 & 4.47 & 7.660 & 8.545 & 0.060 & 0.218 & 28.6$\pm$1.9\\
Theia 546 & 292.2 & -8.5150 & -0.3320 & 0.0621 & 9.59 & 235.09 & -0.17 & 8.013 & 9.279 & 0.081 & 0.067 & 26.5$\pm$2.3\\
Theia 651 & 689.2 & -8.3291 & -0.6812 & 0.1127 & 7.71 & 241.32 & 0.60 & 7.841 & 9.841 & 0.113 & 0.126 & 26.4$\pm$3.5\\
Theia 665 & 274.8 & -8.3901 & -0.7105 & 0.0116 & -10.28 & 236.03 & 3.04 & 8.290 & 9.028 & 0.054 & 0.055 & 26.6$\pm$1.3\\
Theia 708 & 438.9 & -8.0818 & 0.4316 & 0.1524 & 14.09 & 222.57 & -2.85 & 7.245 & 8.404 & 0.074 & 0.168 & 29.9$\pm$2.5\\
Theia 710 & 1024.1 & -8.2479 & 0.3743 & 0.0522 & 8.80 & 235.04 & -6.94 & 8.240 & 8.603 & 0.022 & 0.137 & 27.7$\pm$0.6\\
Theia 713 & 327.5 & -8.3624 & 0.4009 & 0.0674 & 35.24 & 236.59 & -3.38 & 7.876 & 9.325 & 0.097 & 0.097 & 26.9$\pm$2.9\\
Theia 743 & 316.7 & -7.8639 & -0.5915 & 0.0291 & -33.52 & 239.70 & 2.51 & 7.710 & 8.669 & 0.065 & 0.050 & 28.4$\pm$2.0\\
Theia 766 & 699.1 & -7.4536 & -0.0725 & -0.0686 & 1.22 & 232.52 & 3.07 & 7.206 & 7.820 & 0.043 & 0.088 & 31.0$\pm$1.2\\
Theia 771 & 326.0 & -7.6321 & -0.6736 & 0.0704 & -42.49 & 219.84 & 5.86 & 6.690 & 8.083 & 0.094 & 0.115 & 32.0$\pm$3.4\\
Theia 773 & 369.8 & -8.2883 & -0.7817 & -0.0237 & -16.03 & 243.43 & 17.03 & 8.269 & 9.248 & 0.076 & 0.351 & 26.4$\pm$1.9\\
Theia 844 & 646.2 & -8.0261 & 0.6434 & 0.0215 & 31.97 & 249.86 & -12.29 & 7.954 & 9.717 & 0.100 & 0.241 & 26.1$\pm$3.1\\
Theia 862 & 258.0 & -8.6236 & 0.4822 & -0.0216 & 28.80 & 223.27 & -1.75 & 7.812 & 9.155 & 0.084 & 0.039 & 27.1$\pm$2.7\\
Theia 873 & 344.4 & -8.3142 & 0.8120 & 0.0666 & 14.42 & 228.21 & -5.01 & 7.861 & 8.684 & 0.055 & 0.117 & 28.2$\pm$1.5\\
Theia 874 & 319.8 & -8.8318 & 0.0499 & 0.0672 & 19.69 & 222.79 & 2.33 & 7.793 & 9.374 & 0.096 & 0.081 & 27.2$\pm$3.1\\
Theia 900 & 332.6 & -8.7465 & 0.6163 & 0.0009 & 7.44 & 225.72 & 2.48 & 8.100 & 9.124 & 0.062 & 0.046 & 27.2$\pm$1.9\\
Theia 1024 & 525.7 & -8.2050 & -0.4943 & 0.0623 & -8.81 & 225.43 & 8.47 & 7.585 & 8.655 & 0.072 & 0.164 & 28.7$\pm$2.1\\
Theia 1040 & 595.9 & -8.6590 & 0.3035 & 0.0904 & 20.02 & 231.42 & -8.49 & 8.207 & 9.260 & 0.061 & 0.196 & 26.5$\pm$1.7\\
Theia 1117 & 323.5 & -8.1074 & 0.4432 & 0.0034 & 31.86 & 230.70 & -6.38 & 7.608 & 8.847 & 0.076 & 0.113 & 28.4$\pm$2.6\\
Theia 1147 & 664.8 & -8.4180 & -0.5242 & -0.0146 & 9.30 & 239.08 & -5.79 & 7.931 & 9.717 & 0.102 & 0.111 & 26.4$\pm$3.1\\
Theia 1179 & 256.0 & -8.2760 & -0.9456 & -0.0272 & 5.35 & 249.77 & 8.59 & 8.031 & 10.107 & 0.149 & 0.179 & 24.6$\pm$4.0\\
Theia 1188 & 570.8 & -8.1262 & 0.2965 & -0.1126 & 26.79 & 250.05 & 6.31 & 7.941 & 9.874 & 0.109 & 0.180 & 26.0$\pm$3.4\\
Theia 1203 & 278.4 & -7.7747 & -0.1365 & -0.1278 & -30.59 & 217.02 & -6.60 & 6.461 & 8.237 & 0.121 & 0.164 & 31.9$\pm$4.7\\
Theia 1224 & 352.9 & -7.7258 & -0.3586 & 0.0626 & -37.02 & 238.23 & 0.95 & 7.315 & 8.664 & 0.084 & 0.068 & 29.6$\pm$3.1\\
Theia 1297 & 406.7 & -8.2512 & 0.4985 & 0.1175 & 9.12 & 237.11 & 10.33 & 8.210 & 8.749 & 0.035 & 0.222 & 27.6$\pm$0.8\\
Theia 1408 & 734.5 & -8.6036 & -0.2738 & -0.0842 & -21.02 & 226.11 & -1.76 & 7.934 & 9.036 & 0.065 & 0.090 & 27.6$\pm$1.9\\
Theia 1432 & 1744.9 & -8.6074 & 0.4669 & -0.0519 & 2.27 & 259.77 & 8.57 & 8.576 & 10.946 & 0.121 & 0.198 & 24.0$\pm$3.1\\
Theia 2122 & 444.0 & -7.1086 & -0.0639 & 0.0528 & -3.80 & 255.20 & -8.38 & 7.108 & 8.704 & 0.105 & 0.155 & 29.8$\pm$3.6\\
Theia 3263 & 267.9 & -7.3946 & -0.6774 & 0.0437 & 1.96 & 237.47 & 4.34 & 6.992 & 8.465 & 0.103 & 0.083 & 30.1$\pm$3.5\\
Theia 4930 & 250.3 & -8.3646 & -0.8851 & 0.0118 & -27.82 & 219.31 & -2.48 & 7.508 & 8.606 & 0.084 & 0.047 & 29.1$\pm$2.4\\
Theia 5731 & 341.6 & -7.6518 & -0.2318 & -0.1537 & -8.99 & 230.15 & 2.19 & 7.341 & 7.868 & 0.036 & 0.163 & 30.4$\pm$1.0\\
Theia 7593 & 934.1 & -8.1123 & -0.9728 & -0.1692 & 3.51 & 250.90 & 7.44 & 7.683 & 10.331 & 0.147 & 0.256 & 25.6$\pm$4.7\\
Theia 8014 & 345.8 & -7.9405 & 0.5913 & 0.0763 & -11.69 & 227.40 & 1.25 & 6.970 & 8.546 & 0.102 & 0.080 & 30.3$\pm$3.8\\
Trumpler 37 & 285.2 & -8.2202 & 0.5914 & 0.0628 & 37.51 & 246.46 & 6.63 & 8.242 & 9.610 & 0.104 & 0.144 & 25.5$\pm$2.6\\
UBC 56 & 273.7 & -8.9042 & 0.3255 & 0.0309 & 11.44 & 228.73 & 1.48 & 8.440 & 9.208 & 0.048 & 0.041 & 26.3$\pm$1.3\\
UBC 150 & 307.1 & -8.0722 & 0.8125 & -0.1457 & 41.40 & 228.18 & 1.47 & 7.592 & 8.785 & 0.073 & 0.156 & 28.6$\pm$2.5\\
UBC 255 & 974.2 & -8.0166 & -0.8921 & -0.1785 & -11.29 & 257.41 & 0.16 & 7.911 & 10.503 & 0.141 & 0.209 & 25.2$\pm$4.3\\
UPK 17 & 251.3 & -7.4965 & 0.2046 & -0.0278 & -1.67 & 219.18 & -4.65 & 6.457 & 7.778 & 0.097 & 0.075 & 33.4$\pm$3.3\\
UPK 18 & 349.6 & -7.4120 & 0.2353 & -0.0480 & -5.84 & 247.33 & 0.91 & 7.316 & 8.495 & 0.075 & 0.056 & 29.8$\pm$2.6\\
UPK 21 & 467.1 & -7.5782 & 0.1914 & -0.0505 & -15.29 & 247.60 & -7.75 & 7.363 & 8.772 & 0.087 & 0.138 & 29.0$\pm$3.1\\
UPK 24 & 470.7 & -7.6622 & 0.1863 & -0.0190 & 24.13 & 223.42 & 5.00 & 6.799 & 8.227 & 0.101 & 0.085 & 30.8$\pm$3.6\\
UPK 27 & 701.5 & -7.3206 & 0.3698 & 0.0890 & -11.71 & 230.60 & -7.19 & 6.647 & 7.913 & 0.087 & 0.156 & 32.2$\pm$3.1\\
UPK 31 & 425.1 & -7.6468 & 0.2395 & -0.0417 & -3.57 & 240.00 & 11.58 & 7.523 & 8.268 & 0.047 & 0.205 & 29.6$\pm$1.4\\
UPK 80 & 262.6 & -7.7134 & 0.7931 & 0.0783 & 5.68 & 230.14 & 0.36 & 7.125 & 8.201 & 0.070 & 0.080 & 30.6$\pm$2.4\\
UPK 93 & 563.4 & -7.8466 & 0.6546 & 0.0175 & 41.97 & 251.12 & -4.21 & 7.712 & 9.843 & 0.122 & 0.080 & 26.5$\pm$3.9\\
UPK 94 & 272.1 & -7.7628 & 0.8635 & -0.0588 & -0.41 & 238.00 & 3.45 & 7.258 & 8.599 & 0.085 & 0.090 & 29.4$\pm$3.2\\
UPK 119 & 321.1 & -8.0132 & 0.7627 & 0.0458 & 16.76 & 222.68 & 4.40 & 7.253 & 8.314 & 0.071 & 0.084 & 29.6$\pm$2.4\\
UPK 131 & 347.4 & -8.0674 & 0.9492 & 0.0096 & 8.19 & 232.85 & -4.93 & 7.575 & 8.636 & 0.081 & 0.088 & 28.8$\pm$2.3\\
UPK 296 & 378.6 & -8.5265 & 0.3135 & -0.0666 & 10.89 & 232.88 & -4.54 & 8.348 & 8.850 & 0.031 & 0.104 & 27.0$\pm$0.8\\
UPK 333 & 253.5 & -8.7571 & 0.2854 & -0.0491 & -7.76 & 232.46 & -3.91 & 8.311 & 9.295 & 0.058 & 0.086 & 26.8$\pm$1.9\\
UPK 379 & 465.5 & -8.8657 & -0.0230 & 0.1113 & 4.65 & 233.25 & -1.13 & 8.620 & 9.398 & 0.045 & 0.118 & 25.8$\pm$1.1\\
UPK 431 & 425.3 & -8.6842 & -0.4335 & 0.0659 & 2.85 & 245.14 & 2.33 & 8.692 & 9.920 & 0.097 & 0.086 & 24.8$\pm$2.2\\
UPK 448 & 293.9 & -8.7723 & -0.5977 & -0.0959 & -12.84 & 231.74 & -4.82 & 8.533 & 9.108 & 0.038 & 0.132 & 26.3$\pm$0.9\\
UPK 492 & 472.6 & -8.3536 & -0.6631 & 0.0103 & -25.30 & 221.61 & -3.63 & 7.559 & 8.699 & 0.078 & 0.066 & 28.7$\pm$2.3\\
UPK 549 & 368.0 & -8.0049 & -0.9860 & 0.0327 & 7.58 & 249.76 & 1.33 & 7.523 & 9.912 & 0.157 & 0.045 & 26.4$\pm$4.8\\
UPK 560 & 269.6 & -8.0356 & -0.5716 & 0.1407 & -20.66 & 228.77 & 2.25 & 7.721 & 8.247 & 0.048 & 0.148 & 29.2$\pm$1.0\\
UPK 562 & 505.3 & -7.8651 & -0.7607 & 0.0599 & -33.79 & 232.63 & 6.89 & 7.685 & 8.276 & 0.037 & 0.127 & 29.2$\pm$1.0\\
UPK 567 & 410.9 & -7.8962 & -0.5838 & -0.0634 & -1.05 & 233.37 & -1.85 & 7.513 & 8.656 & 0.077 & 0.074 & 29.0$\pm$2.4\\
UPK 578 & 551.8 & -7.6249 & -0.7650 & 0.1073 & -35.09 & 215.17 & -11.61 & 6.545 & 7.905 & 0.098 & 0.236 & 32.3$\pm$3.4\\
UPK 639 & 476.5 & -7.4661 & -0.2031 & -0.0529 & -17.63 & 230.17 & -9.32 & 7.044 & 7.810 & 0.052 & 0.154 & 31.6$\pm$1.5\\
UPK 644 & 326.5 & -7.4202 & -0.1335 & 0.1533 & -12.57 & 232.26 & 6.85 & 7.152 & 7.743 & 0.041 & 0.187 & 31.4$\pm$1.1\\
\end{longtable}

\section{Additional figures}
\label{sec:additional_figures_appendix}

\begin{figure*}
    \includegraphics[scale=1]{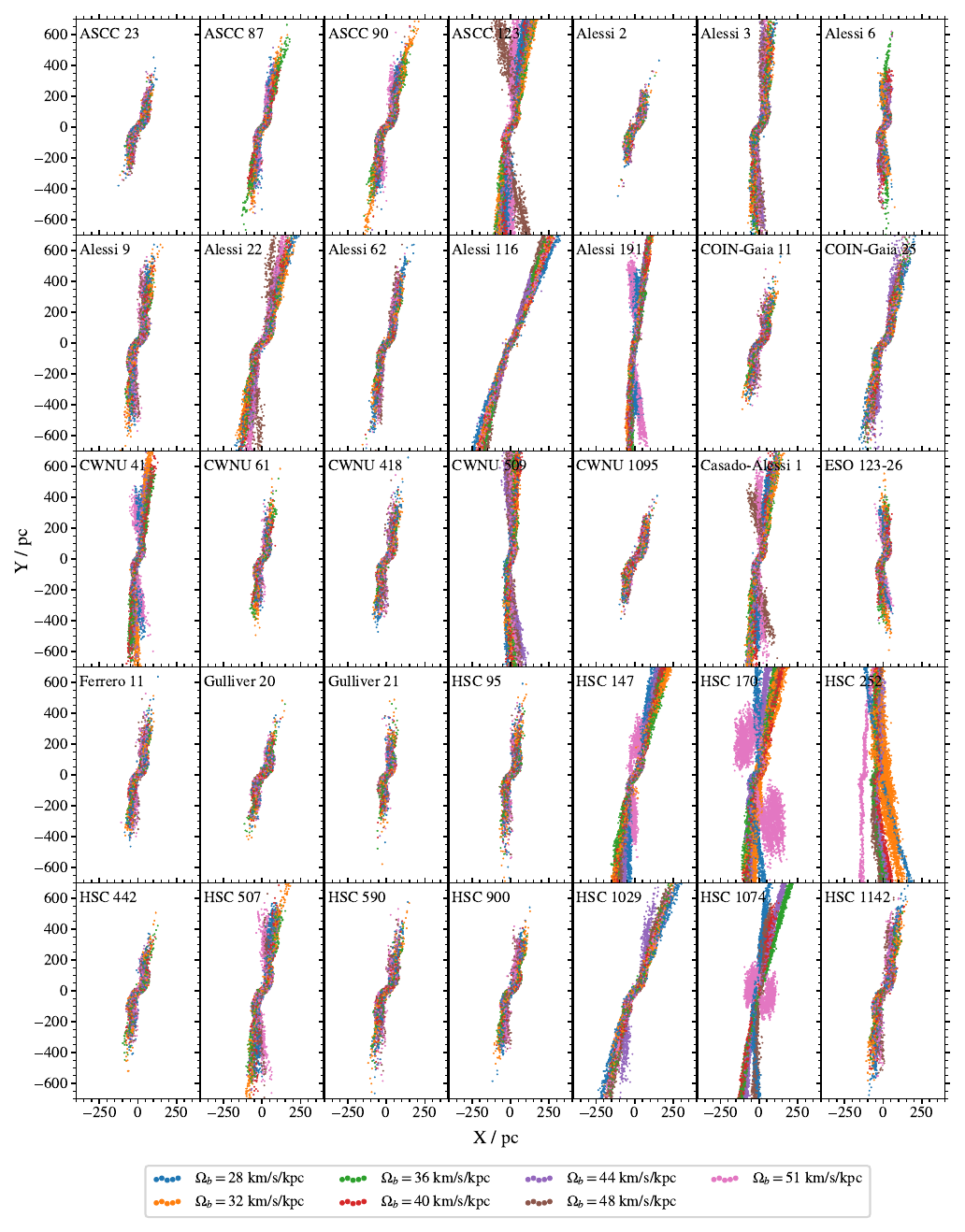}
    \caption{Simulations of cluster dissolution for real open clusters listed in table \ref{tab:clusters}. In each panel we show a cluster's tidal tails in the Galactic plane, for 7 different values of bar pattern speed $\Omega_b$. Stars in each panel are plotted in random order.}
    \label{fig:real_clusters_a}
\end{figure*}

\begin{figure*}
    \includegraphics[scale=1]{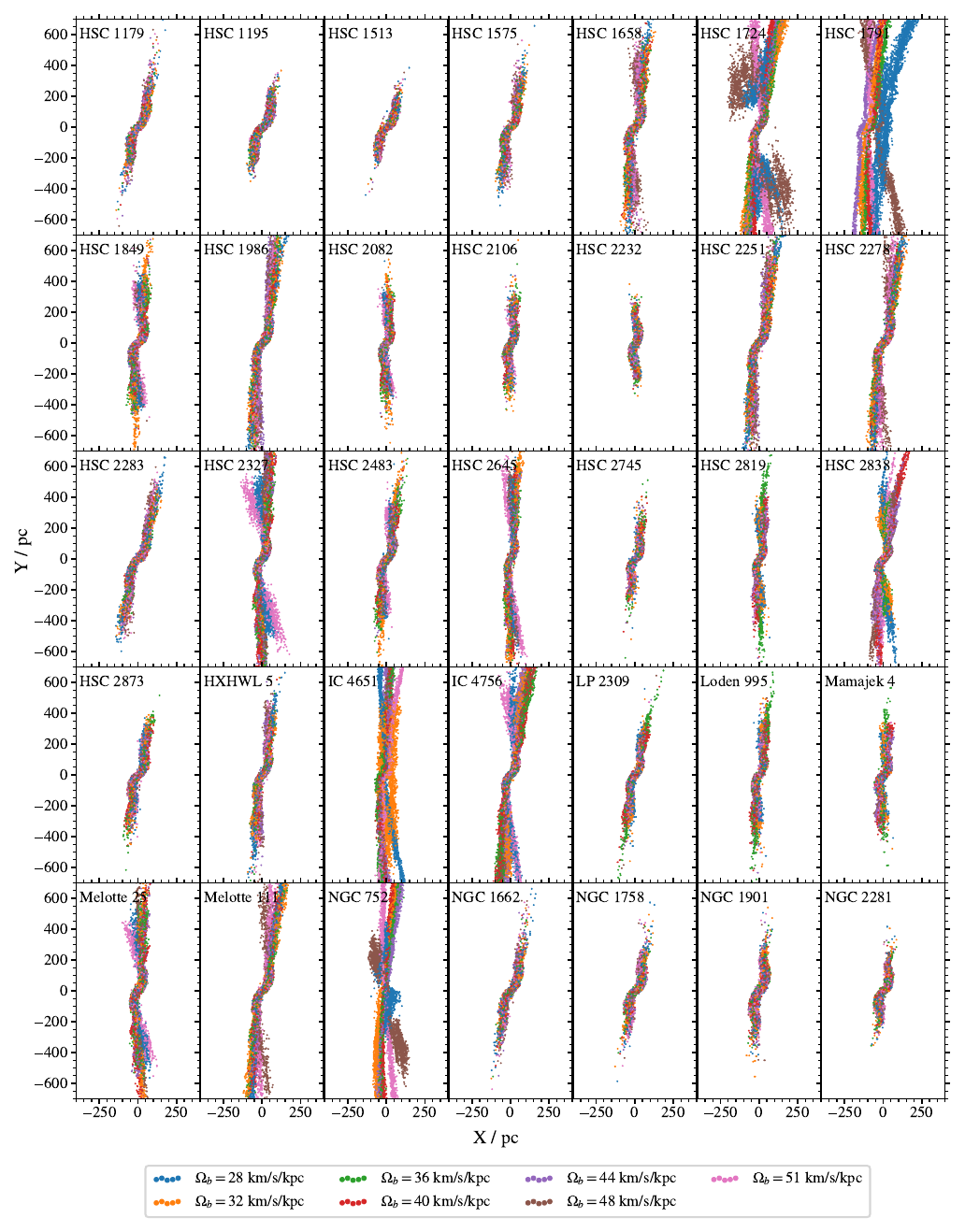}
    \caption{Same as Figure \ref{fig:real_clusters_a} but for the next 35 clusters.}
    \label{fig:real_clusters_b}
\end{figure*}

\begin{figure*}
    \includegraphics[scale=1]{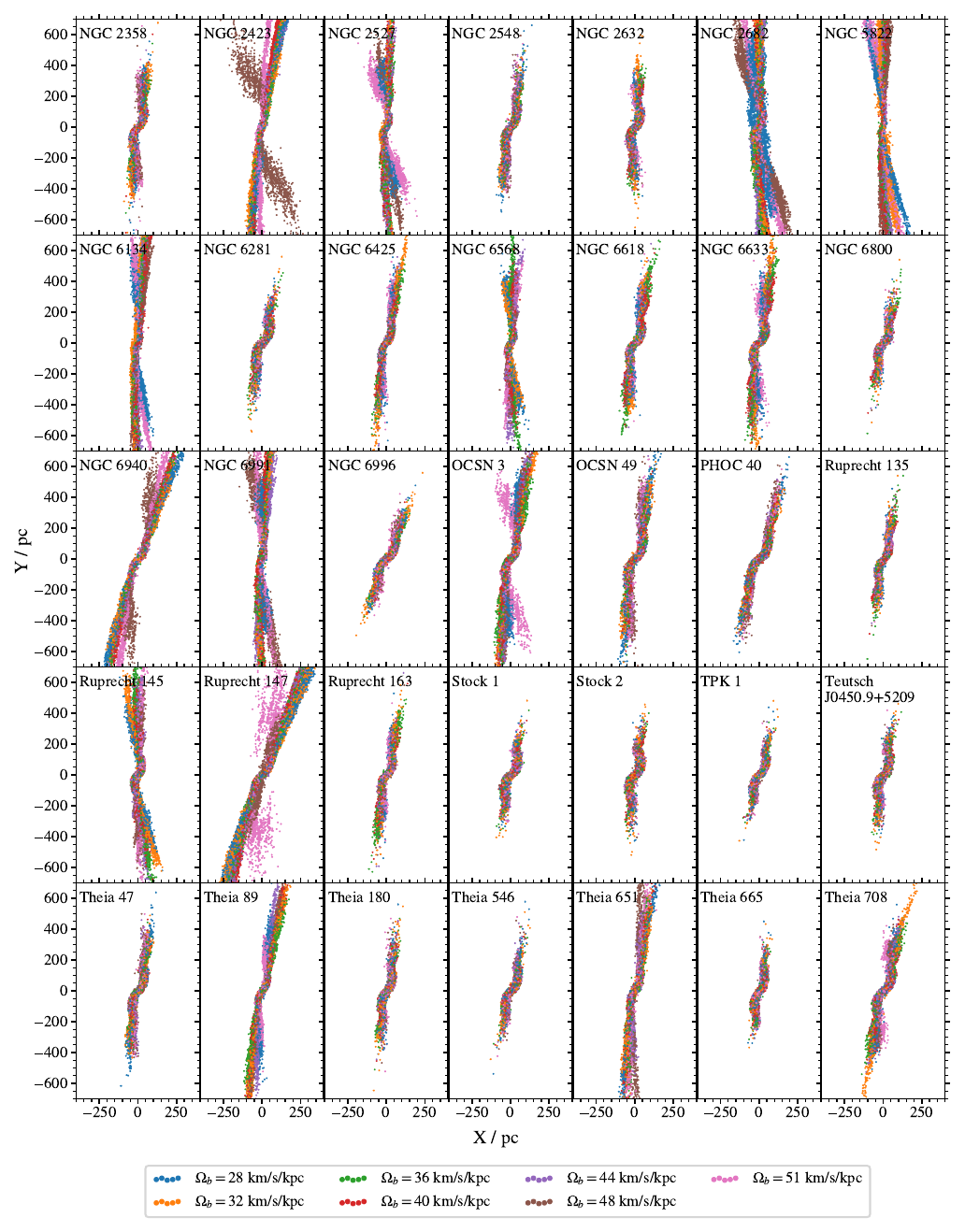}
    \caption{Same as Figure \ref{fig:real_clusters_a} but for the next 35 clusters.}
    \label{fig:real_clusters_c}
\end{figure*}

\begin{figure*}
    \includegraphics[scale=1]{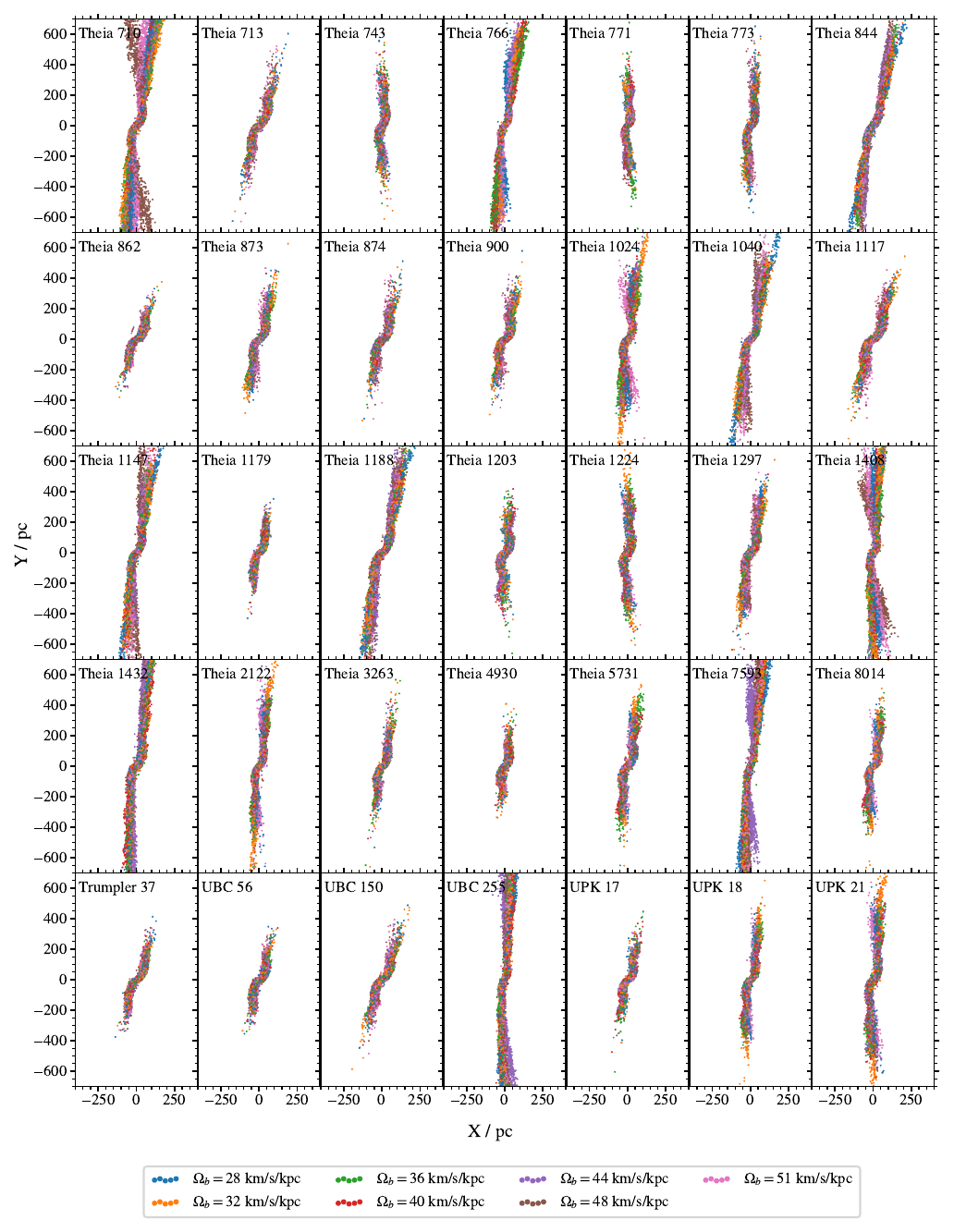}
    \caption{Same as Figure \ref{fig:real_clusters_a} but for the next 35 clusters.}
    \label{fig:real_clusters_d}
\end{figure*}

\begin{figure*}
    \includegraphics[scale=1]{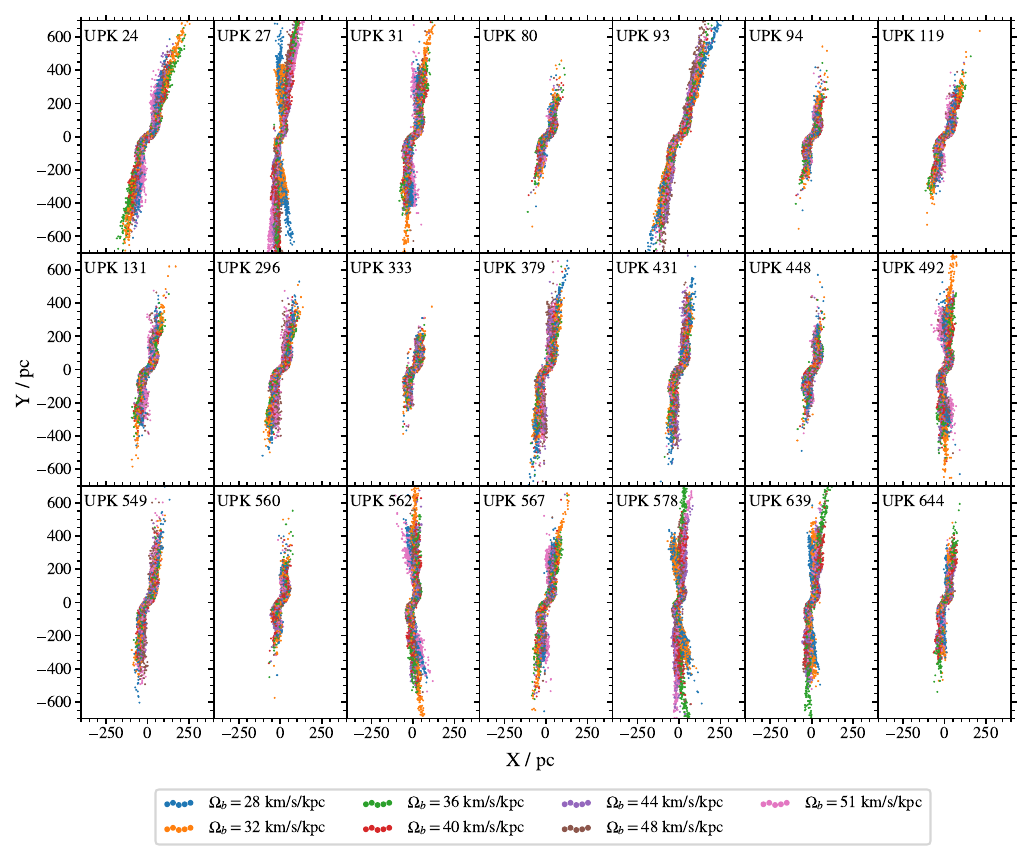}
    \caption{Same as Figure \ref{fig:real_clusters_a} but for the remaining 21 clusters.}
    \label{fig:real_clusters_e}
\end{figure*}

\end{appendix}

\end{document}